\documentclass[PAPER, atlasdraft=false,coverpage=false,UKenglish,cernpreprint, texmf, orcidlogo]{atlasdoc}
\usepackage{atlaspackage}
\usepackage{atlasbiblatex}

\usepackage{atlasphysics}
\usepackage{atlasjetetmiss}
\usepackage{PMGRefs-defs}
\usepackage{ANA-STDM-2020-30-PAPER-defs}
\usepackage{xcolor}
\usepackage{arydshln}
\usepackage{float}
\graphicspath{{logos/}{figures/}}

\usepackage{ANA-STDM-2020-30-PAPER-defs}

\AtlasTitle{Cross-section measurements for the production of a \Wboson in association with high-transverse-momentum jets in $pp$ collisions at $\sqrt{s}$=~13~\TeV with the ATLAS detector}

\AtlasAbstract{%
A set of measurements for the production of a \Wboson in association with high-transverse-momentum jets is presented using 140~fb$^{-1}$ of proton--proton collision data at a centre-of-mass energy of $\sqrt{s}=13$~\TeV collected by the ATLAS detector at the LHC.
The measurements are performed in final states in which the \Wboson decays
into an electron or muon plus a neutrino and is produced in association with jets with $\pt{}>30$~\GeV, where the leading jet has $\pt{}>500$~\GeV.
The angular separation between the lepton and the closest jet with $\pt{}>100$~\GeV
is measured and used to define a collinear phase space, wherein measurements of kinematic properties of the \Wboson and the associated jet are performed.
The collinear phase space is populated by dijet events radiating a \Wboson and events with a \Wboson produced in
association with several jets and it serves as an excellent data sample to probe higher-order theoretical predictions.
Measured differential distributions are compared with predictions from state-of-the-art next-to-leading order multi-leg merged Monte Carlo event generators
and a fixed-order calculation of the \Wplusonejet process computed at next-to-next-to-leading order in the strong coupling constant.
}

\AtlasRefCode{STDM-2020-30}

\PreprintIdNumber{CERN-EP-2024-277}

\AtlasJournalRef{\EPJC 85 (2025) 738}
\AtlasDOI{10.1140/epjc/s10052-025-14078-0}

\AtlasCoverEgroupAnalysisTeam{atlas-STDM-2020-30-analysis-team@cern.ch}

\hypersetup{pdftitle={ATLAS document},pdfauthor={The ATLAS Collaboration}}

\begin{document}

\maketitle

\tableofcontents

\clearpage

%
\newcommand{\AtlasCoordFootnote}{%
ATLAS uses a right-handed coordinate system with its origin at the nominal interaction point (IP)
in the centre of the detector and the \(z\)-axis along the beam pipe.
The \(x\)-axis points from the IP to the centre of the LHC ring,
and the \(y\)-axis points upwards.
Polar coordinates \((r,\phi)\) are used in the transverse plane,
\(\phi\) being the azimuthal angle around the \(z\)-axis.
The pseudorapidity is defined in terms of the polar angle \(\theta\) as \(\eta = -\ln \tan(\theta/2)\) and is equal to the rapidity
$ y = \frac{1}{2} \ln \left( \frac{E + p_z}{E - p_z} \right) $ in the relativistic limit.
Angular distance is measured in units of \(\Delta R \equiv \sqrt{(\Delta y)^{2} + (\Delta\phi)^{2}}\).}

\section{Introduction}

Measurements of electroweak (EW) bosons in the electron or muon decay modes
provide a clean experimental signature to test the EW sector of the Standard Model (SM)
and perturbative quantum chromodynamics (QCD) in high jet-multiplicity final states.
The large production \xsec for a single \Wboson provides a very large data sample
with which to explore the modelling of kinematic variables in the high-momentum phase space.
Moreover, the \Wjets process represents a significant irreducible background
to a wide range of analyses within the Large Hadron Collider (LHC) physics programme, such as
Higgs boson~\cite{HIGG-2018-52,HIGG-2017-14,CMS-HIG-22-006,CMS-HIG-22-008} and top quark measurements~\cite{TOPQ-2020-02,TOPQ-2016-10,CMS-TOP-22-009,CMS-TOP-18-006},
and searches for new physics phenomena beyond the Standard Model (BSM)~\cite{ATLAS:2021twp,ATLAS:2019lng,ATLAS:2020xzu,CMS-SUS-21-003,CMS-SUS-17-007,CMS-SUS-21-007}.
Therefore, an accurate description of this process across a large energy range is critical to the success of the LHC physics programme.

\begin{figure}[!b]
\begin{center}
\subfloat[]{\includegraphics[width=0.35\textwidth]{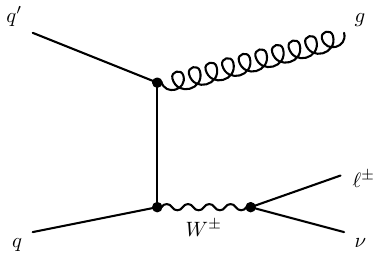}\label{fig:feynman:LO}}\hspace{5em}
\subfloat[]{\includegraphics[width=0.35\textwidth]{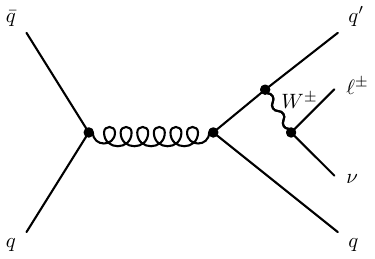}\label{fig:feynman:NLO}}
\caption{
Representative Feynman diagrams for the production of a \Wboson in association with jets.
The \(t\)-channel diagram shown in (a) typically leads to large angular separation between the recoiling jet and the lepton,
while the \(s\)-channel diagram shown in (b) can lead to small angular separations between the lepton and the outgoing quark.
}
\label{fig:feynman}
\end{center}
\end{figure}

At the energy scales being explored at the LHC,
the production of a \Wboson in association with jets is a complex process.
For $W$+1-jet production, at leading order (LO) in the strong coupling constant $\alphas$ the \Wboson recoils against a quark or gluon,
resulting in an experimental signature of a single high-momentum jet balancing the \Wboson decay products (\Cref{fig:feynman:LO}).
At next-to-leading order (NLO) in $\alphas$, new partonic scattering topologies emerge
that can contribute very large enhancements to the overall production rate (\Cref{fig:feynman:NLO}).
Of particular importance are diagrams where dijet production is accompanied by the emission of a real \Wboson
from the incoming or outgoing quarks, as shown in Figure~\ref{fig:feynman}.
These contributions lead to large enhancements in the production rate that are proportional to
$\alphas \ln^2 p_{\text{T}}^{\text{closest jet}} / m_{W}$,
where $p_{\text{T}}^{\text{closest jet}}$ is the transverse momentum of the closest jet,
and $m_W$ is the \Wboson mass.
This results in an overall \xsec enhancement at small values in the distribution of the angular separation between the \Wboson and the closest jet in events with high-\pt{} jets,
referred to as \emph{collinear enhancement}.
The collinear region also contains contributions where the \Wboson is produced in association with larger numbers of jets.
An accurate description of \Wbosons produced in association with many jets demands
theoretical predictions computed at the highest possible order in both the strong and electroweak couplings.

Measurements of \Wbosons emitted from high-momentum partons in the muon channel only
were reported in proton--proton collisions at a centre-of-mass energy of 8~\TeV
by the ATLAS Collaboration using a data sample with 20.3~fb$^{-1}$~\cite{ATLAS:2016jbu} of integrated luminosity,
and a more inclusive measurement by the CMS Collaboration using a data sample with 19.6~fb$^{-1}$~\cite{CMS-SMP-14-023}.
CMS performed an initial measurement of \Wjets events in a high-momentum region using 2.2~fb$^{-1}$ of proton--proton collision data at a centre of mass energy of 13~\TeV in the muon decay channel~\cite{CMS-SMP-16-005}.
This paper presents an extension of those measurements using the full Run~2 data sample corresponding to 140~fb$^{-1}$
of proton--proton collision data at a centre-of-mass energy of 13~\TeV, via a combination of the \xsec measurements from the electron and muon channels.
The large data sample allows the measurement of kinematic variables typically probed in searches for new phenomena beyond the 1~\TeV energy scale.
ATLAS also performed a measurement of \Zjets events in the high-momentum region with the full Run~2 data sample~\cite{ATLAS:2022nrp}.
The measurement presented here is complementary to the \Zjets measurement, and is able to probe higher energy
regimes due to the nearly ten times larger production \xsec.

This paper presents inclusive and differential \xsec measurements of a \Wboson
produced in association with high-$\pt$ jets.
It targets events where the \Wboson decays into a lepton (electron or muon) in association with jets with $\pt{}>30$~\GeV and $|\eta| < 2.5$, where the leading jet has $\pt{}>500$~\GeV.
This phase space is referred to as the \textit{inclusive} region, and measurements of the differential
\xsec for observables sensitive to collinear enhancement in the production rate are performed:

\begin{itemize}
\item \minDRLepJet: angular separation\footnote{\AtlasCoordFootnote} between the lepton and closest jet with transverse momentum greater than 100~\GeV.
Events with $\minDRLepJet>2.6$ (for each jet $i$ in the event) are expected to be dominated by leading-order \Wplusonejet production and referred to as the \textit{back-to-back} region.
The opposing selection $\minDRLepJet<2.6$, expected to be dominated by \(W\)+$\geq2$-jet processes, is referred to as the \textit{collinear} region.
\item \wptratio: the ratio of the transverse momentum of the lepton--neutrino
system (\pTW) to the transverse momentum of the jet that is closest to the lepton.
Values around one represent events where the \Wboson and closest jet have similar momentum,
and therefore are expected to be dominated by leading-order \Wplusonejet production.
Values away from one are expected to be dominated by \(W\)+$\geq2$-jets production.
\item \mjj: the invariant mass of the highest momentum pair of jets.
This observable is measured in the \textit{inclusive 2-jet} region where at least two jets are required.
This is an observable often explored in EW induced measurements~\cite{ATLAS:2020nzk} or BSM searches~\cite{ATLAS:2015gvj,ATLAS:2015eiz,ATLAS:2022yvh},
but is difficult to model in the multi-\TeV kinematic range.
\end{itemize}

In addition, differential measurements are performed of observables sensitive to \Wboson production in association with many jets in the \textit{collinear} region.
These observables are:

\begin{itemize}
\item the transverse momentum of the leading jet,
\item the transverse momentum of the lepton--neutrino system ($p_\text{T}^{\ell\nu}$),
\item the jet multiplicity for jets with $\pT>30~\GeV$, and
\item the total scalar sum of the transverse momentum of all jets (\HT).
\end{itemize}

State-of-the-art event generator programs that are multi-leg matrix element accurate to
NLO in $\alphas$ merged with parton-shower models to emulate
the higher missing orders are compared with the measured data.
In addition, fixed-order predictions for \Wplusonejet production computed at next-to-next-to-leading order (NNLO) in $\alphas$ using the MCFM program~\cite{Campbell:2019dru},
are compared with the measured data.
The data from this publication will be publicly provided through both \textsc{HEPData}~\cite{Maguire:2017ypu} and \textsc{Rivet}~\cite{Bierlich:2019rhm}
and can be used to improve the theoretical description of these high-energy events.

The paper is organised as follows.
The ATLAS detector is described in~\Cref{sec:detector}.
\Cref{sec:samples} presents the data sample and simulated samples used in the measurement,
while~\Cref{sec:objects} details the reconstruction of leptons, jets, and the missing transverse momentum.
The event selection is described in~\Cref{sec:event_selection},
and the estimation of background processes to the measurement are explained in~\Cref{sec:backgrounds}.
The unfolding of detector effects is described in~\Cref{sec:unfolding},
and systematic uncertainties in~\Cref{sec:syst}.
The final results are presented in~\Cref{sec:results}, with concluding remarks in~\Cref{sec:conclusions}.


\section{ATLAS detector \label{sec:detector}}

The ATLAS detector~\cite{PERF-2007-01} at the LHC covers nearly the entire solid angle around the collision point.
It consists of an inner tracking detector surrounded by a thin superconducting solenoid, electromagnetic and hadronic calorimeters,
and a muon spectrometer incorporating three large superconducting air-core toroidal magnets.

The inner-detector system (ID) is immersed in a \qty{2}{\tesla} axial magnetic field
and provides charged-particle tracking in the range \(|\eta| < 2.5\).
The high-granularity silicon pixel detector covers the vertex region and typically provides four measurements per track,
the first hit generally being in the insertable B-layer (IBL) installed before Run~2~\cite{ATLAS-TDR-19,PIX-2018-001}.
It is followed by the SemiConductor Tracker (SCT), which usually provides eight measurements per track.
These silicon detectors are complemented by the transition radiation tracker (TRT),
which enables radially extended track reconstruction up to \(|\eta| = 2.0\).
The TRT also provides electron identification information
based on the fraction of hits (typically 30 in total) above a higher energy-deposit threshold corresponding to transition radiation.

The calorimeter system covers the pseudorapidity range \(|\eta| < 4.9\).
Within the region \(|\eta|< 3.2\), electromagnetic calorimetry is provided by barrel and
endcap high-granularity lead/liquid-argon (LAr) calorimeters,
with an additional thin LAr presampler covering \(|\eta| < 1.8\)
to correct for energy loss in material upstream of the calorimeters.
Hadronic calorimetry is provided by the steel/scintillator-tile calorimeter,
segmented into three barrel structures within \(|\eta| < 1.7\), and two copper/LAr hadronic endcap calorimeters.
The solid angle coverage is completed with forward copper/LAr and tungsten/LAr calorimeter modules
optimised for electromagnetic and hadronic energy measurements respectively.

The muon spectrometer (MS) comprises separate trigger and
high-precision tracking chambers measuring the deflection of muons in a magnetic field generated by the superconducting air-core toroidal magnets.
The field integral of the toroids ranges between \num{2.0} and \qty{6.0}{\tesla\metre}
across most of the detector.
Three layers of precision chambers, each consisting of layers of monitored drift tubes, cover the region \(|\eta| < 2.7\),
complemented by cathode-strip chambers in the forward region, where the background is highest.
The muon trigger system covers the range \(|\eta| < 2.4\) with resistive-plate chambers in the barrel, and thin-gap chambers in the endcap regions.

The luminosity is measured mainly by the LUCID--2~\cite{LUCID2} detector that records Cherenkov light produced in the quartz windows of photomultipliers located close to the beampipe.

Events are selected by the first-level trigger system implemented in custom hardware,
followed by selections made by algorithms implemented in software in the high-level trigger~\cite{TRIG-2016-01}.
The first-level trigger accepts events from the \qty{40}{\MHz} bunch crossings at a rate below \qty{100}{\kHz},
which the high-level trigger further reduces in order to record complete events to disk at about \qty{1}{\kHz}.

A software suite~\cite{SOFT-2022-02} is used in data simulation, in the reconstruction
and analysis of real and simulated data, in detector operations, and in the trigger and data acquisition
systems of the experiment.



\section{Data and simulated event samples}
\label{sec:samples}

The data used in this analysis were recorded with the ATLAS detector from 2015
to 2018 in proton--proton collisions at $\sqrt{s}=$ 13~\TeV\ and
correspond to a total integrated luminosity of 140~\ifb.
The uncertainty in the combined 2015--2018 integrated luminosity is 0.83\%~\cite{DAPR-2021-01},
obtained using the LUCID-2 detector for the primary luminosity measurements,
complemented by measurements using the inner detector and calorimeters.
The mean number of proton--proton interactions per bunch crossing,
including hard scattering events and other interactions in the same and neighbouring bunch crossings (pile-up), was $\langle \mu \rangle = 34$.

Monte Carlo (MC) simulated samples are used to estimate various background processes,
to unfold the data to particle level, to estimate systematic uncertainties and for the comparison with the unfolded data distributions.
Generated events were processed using a full detector simulation~\cite{SOFT-2010-01} based on \GEANT~\cite{Agostinelli:2002hh}
for the detector response to final-state particles. Simulated events were processed and then reconstructed with the same reconstruction algorithms as used for the data.
To account for pile-up, multiple overlaid inelastic $pp$ collisions were simulated with \Pythia[8.186]~\cite{Sjostrand:2007gs}
using the A3 set of tuned parameters (tune)~\cite{ATL-PHYS-PUB-2016-017} and the \NNPDF[2.3lo] set of parton distribution functions (PDFs)~\cite{Ball:2012cx}.
The simulated events are reweighted so that the mean number of proton--proton interactions in simulation matches that observed in each of the data-taking periods.
A summary of the simulated samples is provided in Table~\ref{tab:MCconfig}.
All simulated samples are normalised to the \xsec predicted by the generator unless otherwise specified.

The nominal modelling for the \Wjets signal process uses the \SHERPA[2.2.11] event generator~\cite{Hoeche:2009rj,Bothmann:2019yzt,ATLAS:2021yza}.
Matrix elements accurate to NLO in both the strong and EW couplings (virtual corrections only, denoted NLO EW$_{\text{virt}}$) for up to two jets were computed by \textsc{OpenLoops}~\cite{Cascioli:2011va,Denner:2016kdg},
while LO-accurate matrix elements for up to five jets were calculated with the \textsc{Comix}~\cite{Gleisberg:2008fv} library.
The matrix elements were merged with the \SHERPA parton shower following the \MEPSatNLO
prescription~\cite{Catani:2001cc} and used the tune developed by the \SHERPA authors.
The \NNPDF[3.0nnlo] set of PDFs~\cite{Ball:2014uwa} was used.
The default \SHERPA parton shower~\cite{Schumann:2007mg} based on
Catani--Seymour dipoles and the cluster hadronisation model~\cite{Winter:2003tt} was used.
The phase space was statistically enhanced using the $\maxHTpTV{}$ variable using an analytic enhancement technique,
where \HT{} is the scalar sum of the transverse momenta of all outgoing particles and \pTV{} is the transverse momentum of the vector boson.
The NLO EW$_{\text{virt}}$ corrections are available as alternative generator weights
in three different schemes for the combination with the QCD matrix elements:
additive, multiplicative and exponentiated~\cite{ATLAS:2021yza}.
Nominal predictions were obtained without the NLO EW$_{\text{virt}}$ corrections,
but results are also presented with these corrections using the multiplicative scheme
as default and an uncertainty computed from the envelope of all other combination schemes.
Scale and PDF uncertainties were evaluated on-the-fly~\cite{Bothmann:2016nao} using
7-point variations of the factorisation and renormalisation scales in the matrix elements and parton shower
and using the Hessian PDF eigenvector variations for the \NNPDF[3.0nnlo] set of PDFs.
The scales were varied coherently in the matrix element and parton shower by factors of $0.5$ and $2$ relative to the nominal scale, but avoiding combinations where renormalisation and factorisation scales are varied in opposite directions.

A second multijet merged configuration with NLO accuracy in $\alphas$ was provided using \mgamc{} interfaced with \pythia.
Samples were produced using the \mgamc~v2.6.5~\cite{Alwall:2014hca} program to generate matrix elements for vector-boson production
with up to three additional jets in the final state to NLO accuracy in \alphas.
The showering and subsequent hadronisation were performed using \Pythia[8.240]~\cite{Sjostrand:2014zea}
with the A14 tune~\cite{ATL-PHYS-PUB-2014-021}, using the \NNPDF[2.3lo] PDF set with $\alphas=0.130$.
The different jet multiplicities were merged using the FxFx prescription~\cite{Frederix:2012ps}.
The PDF set used for event generation was \NNPDF[3.1luxQED]~\cite{Bertone:2017bme} computed at NNLO with $\alphas=0.118$.
The samples were generated with additional weights for the PDF eigenvector variations as well
as variations of the renormalisation and the factorisation scales~\cite{Andersen:2014efa}.

Electroweak production of $W$+jj (with $W\rightarrow \ell\nu$) and $Z$+jj (with $Z\rightarrow \ell\ell$ or $\nu\nu$) final states
was simulated with \SHERPA[2.2.11] using LO matrix elements with up to two additional parton emissions.
The matrix elements were merged with the \SHERPA parton shower following the \MEPSatLO
prescription~\cite{Catani:2001cc} and used the tune developed by the \SHERPA authors.  The \NNPDF[3.0nnlo] set of
PDFs was employed. The samples were produced using the VBF approximation, which avoids overlap with semileptonic
diboson topologies by requiring a $t$-channel colour-singlet exchange. The EW production of $\ell\nu$jj is treated as a background
and is subtracted from the data before unfolding. Interference between EW and QCD induced vector-boson processes was neglected.

The production of \ttbar{}+jets events was modelled using the \SHERPA[2.2.12] event generator.
In the \SHERPA[2.2.12] samples NLO-accurate matrix elements, for up to one additional jet
and LO-accurate matrix elements, for up to four additional jets, were calculated with the
\textsc{Comix} and \textsc{OpenLoops} libraries.
Aside from this, the same general configuration as the signal \Wjets described above was used.
Single-top quark production (single-top) included \(s\)- and \(t\)-channel diagrams and the associated production of a single top quark with a \Wboson ($tW$).
All single-top processes were modelled using the
\POWHEGBOX[v2]~\cite{Re:2010bp,Nason:2004rx,Frixione:2007vw,Alioli:2010xd}
generator that provides matrix elements at NLO in \alphas{}.
The \NNPDF[3.0nlo] PDF set was used in the matrix element and events were
interfaced with \PYTHIA[8.230] using the A14
tune and the \NNPDF[2.3lo] set of PDFs.
The inclusive \xsec for single-top was normalised to the theory prediction
calculated at NNLO in QCD with next-to-leading logarithmic (NNLL) soft-gluon corrections~\cite{Kidonakis:2011wy,Kidonakis:2010ux,Kidonakis:2010tc,Kidonakis:2021vob}.

The \POWHEGBOX[v2]~\cite{Nason:2004rx,Frixione:2007vw,Alioli:2010xd} generator
was used to simulate diboson (\(WW\), \(WZ\) and \(ZZ\)) production at NLO
accuracy in $\alphas$~\cite{Nason:2013ydw}. The effect of singly resonant
amplitudes and interference effects due to \(Z/\gamma^*\) and
same-flavour lepton combinations in the final state were included, where appropriate.
Events were interfaced with \PYTHIA[8.210]
for the modelling of the parton shower, hadronisation, and underlying
event, with parameters set according to the \AZNLO
tune~\cite{STDM-2012-23}. The \CT[10] PDF set~\cite{Lai:2010vv} was used
for the hard-scattering processes, whereas the \CTEQ[6L1] PDF
set~\cite{Pumplin:2002vw} was used for the parton shower. The \EVTGEN[1.2.0]
program~\cite{Lange:2001uf} was used to decay bottom and charm hadrons.
The factorisation and renormalisation scales were set to the invariant mass of
the boson pair. An invariant mass of \(m_{\ell\ell} > \SI{4}{\GeV}\) was required
at matrix element level for any pair of same-flavour charged leptons.

Multijet production was simulated using \PYTHIA[8.230] with LO matrix elements matched to the parton shower.
The renormalisation and factorisation scales were set to the geometric
mean of the squared transverse masses of the two outgoing particles in the matrix element.
The \NNPDF[2.3lo] PDF set was used in
the matrix element generation, the parton shower, and the simulation of the
multi-parton interactions. The A14
set of tuned parameters was used. Perturbative uncertainties were estimated
through event weights~\cite{Mrenna:2016sih} that encompass variations
of the scales at which $\alphas$ is evaluated in
the initial- and final-state shower as well as the PDF uncertainty in
the shower and the non-singular part of the splitting functions.

All simulated event samples are corrected to compensate for small efficiency differences between data and simulation.
These include correction factors relating to the efficiency of $b$-tagging for $b$-, $c$- and light-jets,
lepton trigger, reconstruction, identification, and isolation, and the efficiency for jet-to-vertex association requirements. Further details on the event reconstruction can be found in~\Cref{sec:objects}.

\begin{table}[!t]
\newcommand{\MaybeCite}[1]{\ifthenelse{\equal{#1}{}}{}{\cite{#1}}}
\newcommand{\SplitRow}[2]{\ifthenelse{\equal{#1}{#2}}{#1}{\begin{tabular}{@{}l} #1 \\ #2 \end{tabular}}}
\newcommand{\SplitRowCite}[2]{\ifthenelse{\equal{#1}{#2}}{\MaybeCite{#1}}{\begin{tabular}{@{}l} \MaybeCite{#1} \\ \MaybeCite{#2} \end{tabular}}}
\newcommand{\Sample}[9]{\SplitRow{#1}{#3} & \SplitRowCite{#2}{#4} & \SplitRow{#5}{#7} & \SplitRowCite{#6}{#8} & #9 \\ }
\centering
\small
\caption{Overview of Monte Carlo simulated samples used to model signal and background processes.
The perturbative accuracy of the QCD calculation in the matrix element is given in the last column. } \label{tab:MCconfig}

\begin{tabular}{l|ll|ll|c}
\toprule
Process             & \begin{tabular}{c} Matrix element  \\ Parton shower \end{tabular}                                                                                                                           & Ref. & \begin{tabular}{c} PDF set \\ Tune \end{tabular} & Ref. & pQCD \\
\midrule
\Wjets              & \Sample{\SHERPA[2.2.11]}{Gleisberg:2008ta,Gleisberg:2008fv,Cascioli:2011va}{\SHERPA[2.2.11]}{Schumann:2007mg,Hoeche:2012yf}{NNPDF30NNLO}{Ball:2014uwa}{\SHERPA default}{}{0-2j@NLO+3-5j@LO}
\Wjets              & \Sample{\MADGRAPH[v2.6.5]}{Alwall:2014hca}{\PYTHIA[8.240]}{Sjostrand:2014zea}{NNPDF3.1luxQED}{Bertone:2017bme}{A14+FxFx}{}{0-3j@NLO}
\midrule
$Z/\gamma^*$ + jets & \Sample{\SHERPA[2.2.11]}{Gleisberg:2008ta,Gleisberg:2008fv,Cascioli:2011va}{\SHERPA[2.2.11]}{Schumann:2007mg,Hoeche:2012yf}{NNPDF30NNLO}{Ball:2014uwa}{\SHERPA default}{}{0-2j@NLO+3-5j@LO}
\cdashline{2-5}
Diboson             & \Sample{\POWHEG[v2]}{Nason:2013ydw,Nason:2004rx,Frixione:2007vw,Alioli:2010xd}{\PYTHIA[8.210]}{Sjostrand:2014zea}{NLO CT10}{Lai:2010vv}{\AZNLO}{STDM-2012-23}{0j@NLO}
\cdashline{2-5}
$t\bar{t}$+jets     & \Sample{\SHERPA[2.2.12]}{Gleisberg:2008ta,Gleisberg:2008fv,Cascioli:2011va}{\SHERPA[2.2.12]}{Schumann:2007mg,Hoeche:2012yf}{NNPDF30NNLO}{Ball:2014uwa}{\SHERPA default}{}{0-1j@NLO+2-3j@LO}
\cdashline{2-5}
$t$ ($s$-channel)   & \Sample{\POWHEG[v2]}{Alioli:2009je,Nason:2004rx,Frixione:2007vw,Alioli:2010xd}{\PYTHIA[8.230]}{Sjostrand:2014zea}{NLO CT10}{Lai:2010vv}{\textsc{Perugia}2012}{Skands:2010ak}{0j@NLO}
\cdashline{2-5}
$t$ ($t$-channel)   & \Sample{\POWHEG[v2]}{Alioli:2009je,Nason:2004rx,Frixione:2007vw,Alioli:2010xd}{\PYTHIA[8.230]}{Sjostrand:2014zea}{NLO CT10f4}{Lai:2010vv}{\textsc{Perugia}2012}{Skands:2010ak}{0j@NLO}
\cdashline{2-5}
$t + W$             & \Sample{\POWHEG[v2]}{Re:2010bp,Nason:2004rx,Frixione:2007vw,Alioli:2010xd}{\PYTHIA[8.230]}{Sjostrand:2014zea}{NLO CT10}{Lai:2010vv}{\textsc{Perugia}2012}{Skands:2010ak}{0j@NLO}
\cdashline{2-5}
EW $W$+jj           & \Sample{\SHERPA[2.2.11]}{Gleisberg:2008ta,Gleisberg:2008fv,Cascioli:2011va}{\SHERPA[2.2.11]}{Schumann:2007mg,Hoeche:2012yf}{NNPDF30NNLO}{Ball:2014uwa}{\SHERPA default}{}{2j@LO}
\cdashline{2-5}
Multijet            & \Sample{\PYTHIA[8.230]}{Sjostrand:2014zea}{\PYTHIA[8.230]}{Sjostrand:2014zea}{\NNPDF[2.3lo]}{Ball:2012cx}{A14}{}{2j@LO}
\bottomrule
\end{tabular}
\end{table}



\section{Event reconstruction}
\label{sec:objects}

Events are required to satisfy a set of quality requirements that ensures the detector was in good operating condition~\cite{DAPR-2018-01}.
In addition, events are required to have a reconstructed primary vertex with two or more associated tracks,
where the primary vertex is chosen as the vertex with the highest scalar sum of the $\pt^2$ of associated tracks~\cite{ATL-PHYS-PUB-2015-026}.
The data sample was collected using either a single-electron or single-muon trigger with different transverse momentum,
isolation and identification criteria, which depended on the data-taking periods,
each characterised by a different instantaneous luminosity~\cite{ATLAS:2019dpa,ATLAS:2020gty,TRIG-2019-03}.

Baseline electrons are reconstructed by algorithms that combine ID tracks and clusters of energy deposits in the electromagnetic calorimeter.
The track associated to the electron candidate is required to have a longitudinal impact parameter ($|z_0 \sin \theta|$) less than 0.5~mm
and the significance of the transverse impact parameter ($|d_0/\sigma_{d_0}|$) is required to be less
then 5.0 to ensure the electron candidate originates from the primary vertex.
Baseline electron candidates are required to have $\pt>\SI{10}{\GeV}$ and $|\eta|< 2.47$.
Additionally, electrons reconstructed in the transition region between the calorimeter barrel and endcap regions,
which contains a relatively large amount of inactive material,
are removed by excluding electrons with $1.37 < |\eta| < 1.52$.
Baseline electron candidates are identified using the electromagnetic shower shape, calorimeter energy to tracker momentum ratio,
and other discriminating variables that are combined into likelihood-based parameters;
they are selected if they satisfy the \textit{TightLLH} quality criterium~\cite{EGAM-2018-01}.
The signal electron candidates must additionally satisfy the \textit{Tight\_VarRad} isolation requirement~\cite{EGAM-2018-01},
which uses the calorimeter isolation and track-based isolation cones around the electron candidate.
Unlike muon candidates defined below, the isolation cones for electrons are variable-radius.
The electron energy scale and resolution is determined using electrons from
\Zboson decays in data and applied to simulation~\cite{EGAM-2021-02}.

Baseline muon candidates are reconstructed by combining tracks from the ID and MS sub-detectors via a global refit.
Baseline muons are required to be of at least \textit{Medium} quality~\cite{MUON-2018-03}
with transverse momentum greater than $\SI{10}{\GeV}$ and $|\eta|<2.4$.
To ensure signal muon candidates are well isolated from other tracks,
they are required to satisfy a combination of calorimeter and track-based isolation criteria
implemented into the \textit{Loose\_FixedRad} isolation working point~\cite{MUON-2018-03}.
The associated ID track is required to satisfy $|z_0 \sin \theta|<0.5$~mm
and $|d_0/\sigma_{d_0}| < 3.0$ to be consistent with the primary vertex.
The muon energy scale and resolution is determined using muons from \Zboson
decays in data and applied to simulation~\cite{MUON-2022-01}.

Jets are clustered using the \antikt algorithm~\cite{Cacciari:2008gp} implemented in the \Fastjet package~\cite{Fastjet} with a radius parameter of $R = 0.4$.
The jets are clustered from particle flow objects~\cite{PERF-2015-09},
which are charged-particle tracks matched to the hard-scatter vertex and calorimeter
energy clusters following an energy subtraction algorithm that removes the calorimeter deposits associated with good-quality tracks from any vertex.
Different MC-based calibration steps are applied to the reconstructed jets~\cite{JETM-2018-05},
including an area-based correction to account for energy contributions from pile-up interactions,
a \pt- and $\eta$-dependent calibration to match the generator-level energy scale of the jets,
and the \enquote{global sequential calibration} to minimize energy calibration differences
between quark- and gluon-initiated jets.
Finally, an in situ calibration is applied to jets in data to match the energy scale in simulation.
All calibrated jets are required to have $\pt > \SI{30}{\GeV}$ and $|y| <2.5$.
Jets with $\pt < \SI{60}{\GeV}$ and $|y| < 2.4$ must also satisfy a requirement
based on the output of the multivariate `jet vertex tagger' (JVT) algorithm,
which is used to identify and reject jets from pile-up vertices~\cite{PERF-2014-03}.
The \textit{Tight} (default) working point is used and jets failing to satisfy the JVT requirements are discarded.
These remaining jets are categorised as baseline jets.

Baseline jets originating from $b$-quarks need to be rejected to suppress background contributions from processes involving top quarks.
Jets containing a $b$-hadron (\bjets) are identified using the DL1r $b$-tagging procedure~\cite{FTAG-2019-07},
a deep learning multivariate algorithm trained using information on tracks and secondary vertices.
Jets are rejected if they satisfy the DL1r working points tuned to have a 60\% efficiency on average for jets
associated with generated $b$-hadrons in simulated \ttbar events.
This working point is chosen to improve the purity of the \ttbar control region described in~\Cref{sec:backgrounds}.
At the chosen working point, the light-jet (charm-jet) rejection measured in
\ttbar MC simulation is about a factor of 1155 (29) on average.

The missing transverse momentum (with magnitude \met) is estimated as the negative vector sum of the transverse momentum
of all identified electrons, photons, jets and muons~\cite{JETM-2020-03}.
All electrons, muons and jets described above are used as input to the reconstruction.
Tracks not associated with any such object are included in the so-called soft term.

An overlap-removal procedure is applied to uniquely identify signal lepton candidates and signal jet candidates in an event.
Baseline electrons are removed if they share an ID track with a baseline muon.
Baseline jets within $\DeltaR < 0.2$ of a baseline electron are removed, then any remaining baseline electrons with $0.2 < \DeltaR < 0.4$ of a baseline jet are removed.
Finally, baseline muons closer than $\DeltaR = 0.4$ to any remaining baseline jets are removed.



\section{Event selection \label{sec:event_selection}}

Events are required to contain a single electron or muon satisfying the criteria described in Section~\ref{sec:objects}
and summarised in the first section of Table~\ref{selection:recoSelection}.
The leading (highest \pt) lepton is required to be matched to the single lepton that triggered the event and have transverse momentum \pt{}$>30$~\GeV.
In addition, the angular separation between the lepton and any selected jet is required to be $\Delta R(\ell,\text{jet})>0.4$.
To enrich the selected data sample in events where the \Wboson is produced in association with many jets,
events are required to contain at least one jet with transverse momentum \pt{}$>$~500~\GeV.
Contributions from \Zjets and di-leptonic \ttbar{} processes are suppressed by imposing a veto on any additional leptons in the event.
To further suppress the contribution of top-quark backgrounds to the measurement region, a \bjet veto is applied.
The event selection is summarised in the second section of Table~\ref{selection:recoSelection}.

The selection described above forms the \textit{inclusive} region.
Three additional regions that are subsets of the inclusive selection are defined.
The first is the \textit{inclusive-2 jets} region, where a requirement of at least two jets is applied.
The second region is the \textit{collinear region}, where the angular separation between the lepton and the closest jet
with $\pt>100$~\GeV is required to be $\minDRLepJet<2.6$.
The final region is the \textit{back-to-back region}, where the angular separation requirement is inverted
relative to the collinear region and so that $\minDRLepJet>2.6$.
A summary of the region selection criteria can be found at the bottom section of Table~\ref{selection:recoSelection}.

\begin{table}[htp]
\begin{center}
\caption{Detector-level selections used to define the signal regions at reconstruction level (see text for details).
\label{selection:recoSelection} }
\begin{tabular}{cc}
\toprule
\multicolumn{2}{c}{\textbf{Object selection}}                                           \\
\midrule
Lepton momentum                        & $p_\text{T}^{\ell} > 10$~\GeV                  \\
Electron pseudorapidity                & $|\eta| < 1.37,\ 1.52 < |\eta| < 2.47$         \\
Muon pseudorapidity                    & $|\eta|<2.4$                                   \\
Jet transverse momentum                & $p_\text{T}^{\text{jet}} > 30$~\GeV            \\
Jet rapidity                           & $|y|<2.5$                                      \\
\midrule
\multicolumn{2}{c}{ \textbf{Event selection}}                                           \\
\midrule
Leading lepton                         & $p_\text{T}^{\ell} > 30$~\GeV, trigger matched \\
Number of leptons                      & $n_{\ell} = 1$                                 \\
Angular separation of leptons and jets & $\Delta R(\ell,\text{jet})>0.4$                \\
Number of jets                         & $n_\text{jet}  \geq 1$                         \\
Leading jet transverse momentum        & $p_\text{T}^{\text{jet}} > 500$~\GeV           \\
\bjet veto                             & $n_{b\text{-jet}}=0$                           \\
\midrule
\multicolumn{2}{c}{\textbf{Signal regions}}                                             \\
\midrule
Inclusive                              & All requirements above                         \\
Inclusive 2-jet                        & Inclusive plus $n_\text{jet}  \geq 2$          \\
Collinear                              & Inclusive plus \minDRLepJet $<$ 2.6            \\
Back-to-back                           & Inclusive plus \minDRLepJet $>$ 2.6            \\
\bottomrule
\end{tabular}
\end{center}
\end{table}



\section{Background estimation}
\label{sec:backgrounds}

Several SM processes can give rise to a signature similar to the \Wjets signal process.
These background processes need to be estimated and subtracted from data before performing the unfolding procedure.
The dominant prompt-lepton backgrounds arise from processes involving the decay of an EW boson to final states with leptons,
and are therefore primarily associated with \ttbar{} and \Zjets processes.
The \ttbar{} and \Zjets{} backgrounds contribute up to 8\% of the total expected event yield across the four regions.
They contribute equally to both the electron and muon channels.
Backgrounds can also arise from multijet events for which the
leptons are either non-prompt (i.e. arising from heavy-quark decays),
or jets mis-identified as leptons, collectively referred to as `fake leptons'.
The latter contribution represents a significant background in the back-to-back region of the electron selection (up to 20\%),
while the contribution of fake leptons in the muon channel is a sub-dominant background.
Multijet backgrounds in all other regions contribute up to 8\% (3\%) for electrons (muons), respectively.
The single-top-quark and diboson processes represent sub-dominant backgrounds.
Their contributions are estimated directly from simulation,
and are expected to account for up to 2\% (3\%) for electrons (muons).
The EW production of $V+$jj contributes up to 10\% of the total expected event yield for large invariant jet mass in the inclusive-2 jets region.

The dominant backgrounds, namely \ttbar, \Zjets, and multijet are estimated by using a semi-data-driven technique separately for each background.
The normalisation of these background contributions is extracted from data in dedicated control regions enriched in their respective processes.
The goal of this method is to extract normalisation factors for each of the backgrounds without introducing a bias from
the contributions of other backgrounds that are known to be mis-modelled in the high momentum region of phase space.
The control regions are constructed such that they are kinematically similar to the inclusive measurement region,
except for a few selections that allow the control region to be enriched in the targeted background process.
These factors are statistically compatible with normalisation factors derived within each individual signal region.
Since the perturbative accuracy of the \ttbar{} and \Zjets{} samples varies with additional QCD emissions,
the normalisation factors are parameterised as a function of the number of jets ($n_{\text{jet}}$).
The multijet normalisation factor is parameterised as a function of lepton \pt and $|\eta|$
to capture the variations of the fake-lepton composition as a function of detector geometry and momentum of the fake leptons.
The \ttbar{} and \Zjets{} normalisation factors are determined first using the MC predictions for multijets, and then applied to
the multijets control region where the multijet normalisation factor is subsequently derived.
This approach is valid because of the small contribution of the multijets process in the \ttbar{} and \Zjets{} control regions.

The \ttbar{} control region is defined in the same way as the inclusive region,
except the \bjet veto is replaced with the requirement of at least two $b$-tagged jets.
The purity of this region in terms of \ttbar{} events is more than 80\%, with small contributions from single-top and \Wjets processes.
In both the electron and muon channels, the extracted normalisation factors are $0.8\pm0.1$ for the 2-jet and 3-jet bins and close to unity for the higher jet multiplicities.
The background contribution to the 1-jet bin is less than 1\% of the total expected event yield for all samples.
Good agreement between the background model with the normalisation factors applied and data is observed as a function of all observables measured.
The only exception is the \mjj{} distribution, where the expected event yield is 1$\sigma$ higher than the data for $\mjj>1$~\TeV,
where $\sigma$ is the total uncertainty. An additional \mjj{} modelling uncertainty is applied to account for this mis-modelling as described in Section~\ref{sec:syst}.

Most of the \Zjets contribution in the signal region consists of events where one of the two leptons
fails the kinematic or geometric requirements, resulting in a single-lepton event.
The remaining \Zjets contributions arise from one of the leptons being mis-identified as a jet.
The \Zjets control region is defined in the same way as the inclusive region,
except the single-lepton requirement is replaced by requiring exactly two same-flavour opposite-sign
leptons with an invariant mass between $60<m_{\ell\ell}<\SI{120}{\GeV}$.
In the \Zjets control region, the purity of the \Zjets process is approximately 90\% with the remaining contributions arising primarily from diboson production.
The average normalisation factor for \Zjets events with up to three jets is $1.05\pm0.05$ and decreases to $0.80\pm0.05$ for events with greater than five jets.
Mis-identified jets that are matched to generator-level electrons from the \Zboson decay
are excluded from the jet-multiplicity-dependent normalisation factor in the signal region
to match the jet counting in the control region, where both electrons are reconstructed.
The corrected background model provides an excellent description of the data in the control region
as a function of all measurement observables, except for \mjj{}.
As for the \ttbar{} process, an additional modelling uncertainty is applied to the \Zjets{} of the \mjj{} distribution in the signal region.

The multijet control region is defined in the same phase space as each of the signal regions,
but the lepton is instead required to fail to satisfy the impact parameter or the isolation lepton criteria.
The purity of this region in terms of multijet events is more than 80\%, with small contributions from top and \Wjets processes.
The normalisation factor is roughly independent of lepton \pt{} for the electron and muon channels,
but ranges from $0.3\pm0.1$ to $0.8\pm0.1$ for high to low electron $|\eta|$.
After the application of the lepton \pt and $|\eta|$ correction factors to the multijet MC sample,
an excellent description of the data is observed for all observables measured.
The modelling of the multijet estimate is additionally checked in a region where the lepton signal criteria are satisfied,
but the missing transverse momentum is below 100~\GeV. This validation region has a small 8\% overlap with the measurement signal region
while the multijet background accounts for more than half of the total event yield.
Good agreement is found between the simulated events and observed data in the validation region for the full set
of measured observables; therefore, no additional uncertainties are assigned to the multijet normalisation factors.

The event yields in each measured signal region, after the corrections extracted from the control regions as described above, are summarised in~\Cref{tab:backgroundContribution} with signal theory uncertainties included.
The \Wjets signal prediction is obtained from the \SHERPA[2.2.11] simulation.
The large $\pt$ requirement on the leading jet selects high jet multiplicity events, so the total uncertainty in the number of \Wjets is dominated by LO matrix elements for \SHERPA[2.2.11].
Overall, the total signal-plus-background predictions are in good agreement with the observed data across all regions for both the electron and muon channels.
Differential distributions in the electron and muon channels are shown in Figure~\ref{fig:backgrounds:reco} for \minDRLepJet, \wptratio, \pTW, and leading jet \pt.
The combined signal plus background predictions are in good agreement with the observed data for the full set of observables,
except for the \mjj distribution where the combined signal plus background prediction is about 1--2$\sigma$ higher than the data including all signal and all background uncertainties,
similar to that observed in the \ttbar{} and \Zjets{} control regions.

\begin{table}[ht]
\sisetup{group-minimum-digits=4}
\centering
\caption{Event yields in the electron and muon channels for the different signal regions
after the application of the background normalisation factors.
The uncertainties in the predictions include statistical and systematic uncertainties added in quadrature.
The nominal \SHERPA[2.2.11] signal prediction is used to compare with the data.
Theoretical uncertainties are also included for signal.}
\begin{tabular}{lr@{\(\,\pm\,\)}rr@{\(\,\pm\,\)}rr@{\(\,\pm\,\)}rr@{\(\,\pm\,\)}r}
\toprule
Process              & \multicolumn{2}{c}{Inclusive}                   & \multicolumn{2}{c}{Collinear}        & \multicolumn{2}{c}{Inclusive 2-jet}          & \multicolumn{2}{c}{Back-to-back}                                                                                 \\
\midrule
& \multicolumn{8}{c}{ \textbf{Electron channel} }                                                                                                                                                                                                          \\
\midrule
\ttbar{}             & \numRF[2]{7229}                                 & \numRF[2]{1659}                      & \numRF[2]{5808}                              & \numRF[2]{1236}                      & \numRF[2]{7073}   & \numRF[2]{1648}  & \numRF[2]{1421}  & \numRF[1]{432}  \\
Multijet             & \numRF[3]{14628}                                & \numRF[2]{1900}                      & \numRF[2]{6344}                              & \numRF[2]{1344}                      & \numRF[3]{11598}  & \numRF[2]{1678}  & \numRF[2]{8283}  & \numRF[1]{732}  \\
Single top           & \numRF[2]{2149}                                 & \numRF[1]{595}                       & \numRF[2]{1608}                              & \numRF[1]{463}                       & \numRF[2]{2056}   & \numRF[1]{578}   & \numRF[2]{541}   & \numRF[2]{135}  \\
EW $V+$jj            & \numRF[2]{4182}                                 & \numRF[1]{812}                       & \numRF[2]{3029}                              & \numRF[1]{590}                       & \numRF[2]{3746}   & \numRF[1]{733}   & \numRF[3]{1153}  & \numRF[2]{222}  \\
\Zjets               & \numRF[3]{6352}                                 & \numRF[2]{209}                       & \numRF[3]{5110}                              & \numRF[2]{199}                       & \numRF[3]{5990}   & \numRF[2]{200}   & \numRF[4]{1241}  & \numRF[2]{68}   \\
$W(\to\tau\nu)$+jets & \numRF[2]{6494}                                 & \numRF[1]{968}                       & \numRF[2]{4034}                              & \numRF[1]{535}                       & \numRF[2]{5619}   & \numRF[1]{819}   & \numRF[2]{2461}  & \numRF[1]{443}  \\
Diboson              & \numRF[3]{3443}                                 & \numRF[2]{341}                       & \numRF[3]{2224}                              & \numRF[2]{237}                       & \numRF[3]{3073}   & \numRF[2]{320}   & \numRF[3]{1219}  & \numRF[2]{106}  \\
\midrule
\Wjets               & \numRF[2]{77339}                                & \numRF[2]{25099}                     & \numRF[2]{53989}                             & \numRF[2]{21196}                     & \numRF[2]{68367}  & \numRF[2]{24403} & \numRF[2]{23350} & \numRF[1]{3980} \\
\midrule
SM Total             & \numRF[3]{121816}                               & \numRF[2]{25294}                     & \numRF[2]{82146}                             & \numRF[2]{21319}                     & \numRF[3]{107522} & \numRF[2]{24576} & \numRF[2]{39670} & \numRF[1]{4112} \\
Data                 & \multicolumn{2}{l}{\numRF[5]{119880}}           & \multicolumn{2}{l}{\numRF[4]{79887}} & \multicolumn{2}{l}{\numRF[5]{105740}}        & \multicolumn{2}{l}{\numRF[4]{39993}}                                                                             \\
\midrule
& \multicolumn{8}{c}{ \textbf{Muon channel} }                                                                                                                                                                                                              \\
\midrule
\ttbar{}             & \numRF[2]{5530}                                 & \numRF[2]{1188}                      & \numRF[2]{4845}                              & \numRF[2]{1000}                      & \numRF[2]{5492}   & \numRF[2]{1186}  & \numRF[2]{685}   & \numRF[2]{196}  \\
Multijet             & \numRF[3]{3637}                                 & \numRF[2]{154}                       & \numRF[3]{2461}                              & \numRF[2]{127}                       & \numRF[3]{3251}   & \numRF[2]{158}   & \numRF[4]{1176}  & \numRF[2]{56}   \\
Single top           & \numRF[2]{1742}                                 & \numRF[1]{479}                       & \numRF[2]{1369}                              & \numRF[1]{391}                       & \numRF[2]{1693}   & \numRF[1]{470}   & \numRF[3]{373}   & \numRF[2]{89}   \\
EW $V+$jj            & \numRF[2]{3572}                                 & \numRF[1]{701}                       & \numRF[2]{2636}                              & \numRF[1]{518}                       & \numRF[2]{3233}   & \numRF[1]{638}   & \numRF[2]{936}   & \numRF[2]{183}  \\
\Zjets               & \numRF[3]{4328}                                 & \numRF[2]{148}                       & \numRF[3]{3585}                              & \numRF[2]{144}                       & \numRF[3]{4149}   & \numRF[2]{143}   & \numRF[3]{743}   & \numRF[2]{44}   \\
$W(\to\tau\nu)$+jets & \numRF[2]{5096}                                 & \numRF[1]{779}                       & \numRF[2]{3321}                              & \numRF[1]{476}                       & \numRF[2]{4484}   & \numRF[1]{680}   & \numRF[3]{1775}  & \numRF[2]{321}  \\
Diboson              & \numRF[3]{2881}                                 & \numRF[2]{299}                       & \numRF[3]{1929}                              & \numRF[2]{210}                       & \numRF[3]{2593}   & \numRF[2]{281}   & \numRF[3]{953}   & \numRF[2]{90}   \\
\midrule
\Wjets               & \numRF[2]{72016}                                & \numRF[2]{23793}                     & \numRF[2]{52024}                             & \numRF[2]{20381}                     & \numRF[2]{64652}  & \numRF[2]{23235} & \numRF[2]{19992} & \numRF[2]{3468} \\
\midrule
SM Total             & \numRF[2]{98802}                                & \numRF[2]{23858}                     & \numRF[2]{72170}                             & \numRF[2]{20426}                     & \numRF[2]{89548}  & \numRF[2]{23296} & \numRF[2]{26632} & \numRF[2]{3498} \\
Data                 & \multicolumn{2}{l}{\ \,\,\!\numRF[4]{94229}}    & \multicolumn{2}{l}{\numRF[4]{67969}} & \multicolumn{2}{l}{\ \,\,\!\numRF[4]{85128}} & \multicolumn{2}{l}{\numRF[4]{26260}}                                                                             \\
\bottomrule
\label{tab:backgroundContribution}
\end{tabular}
\end{table}

\begin{figure}[!htbp]
\graphicspath{{./figures/background/}}
\begin{center}
\subfloat[]{\includegraphics[width=0.45\textwidth]{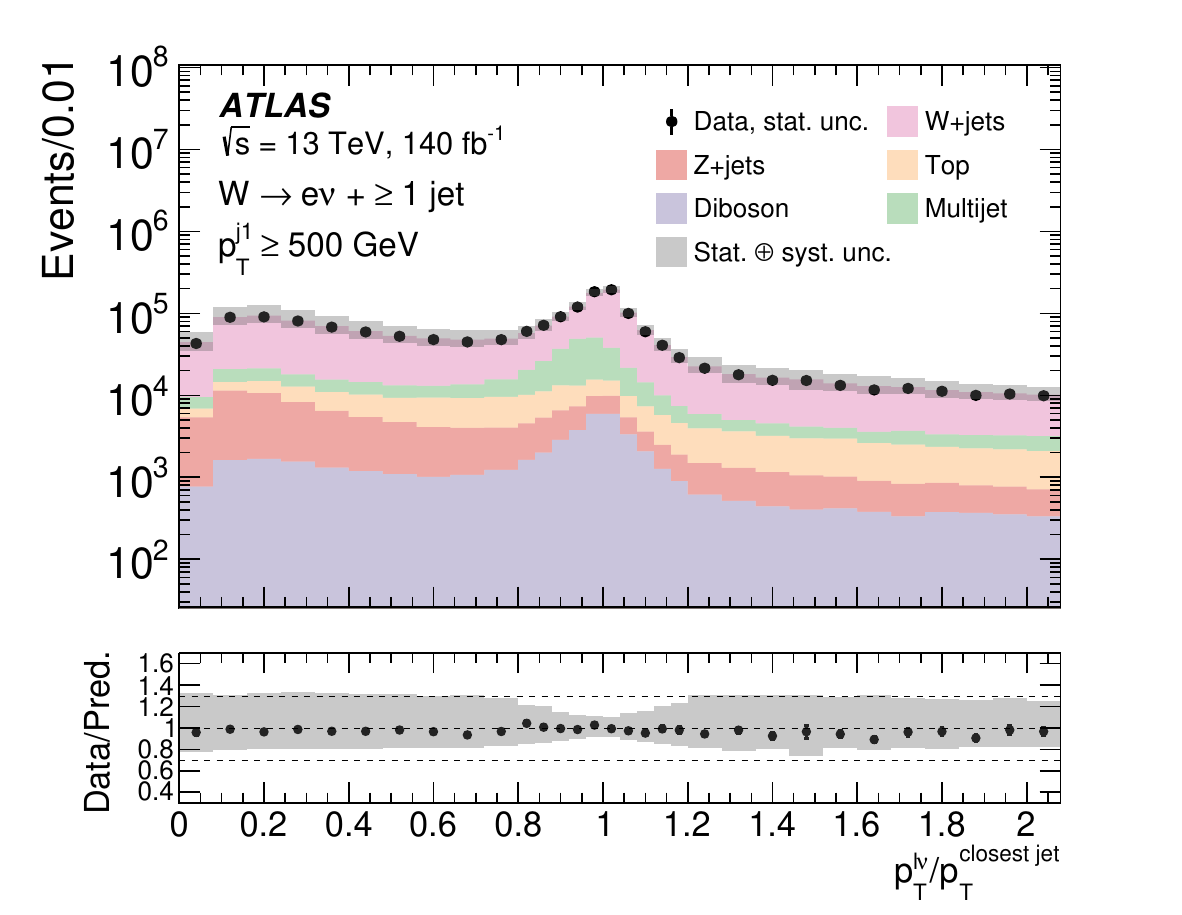}}\qquad
\subfloat[]{\includegraphics[width=0.45\textwidth]{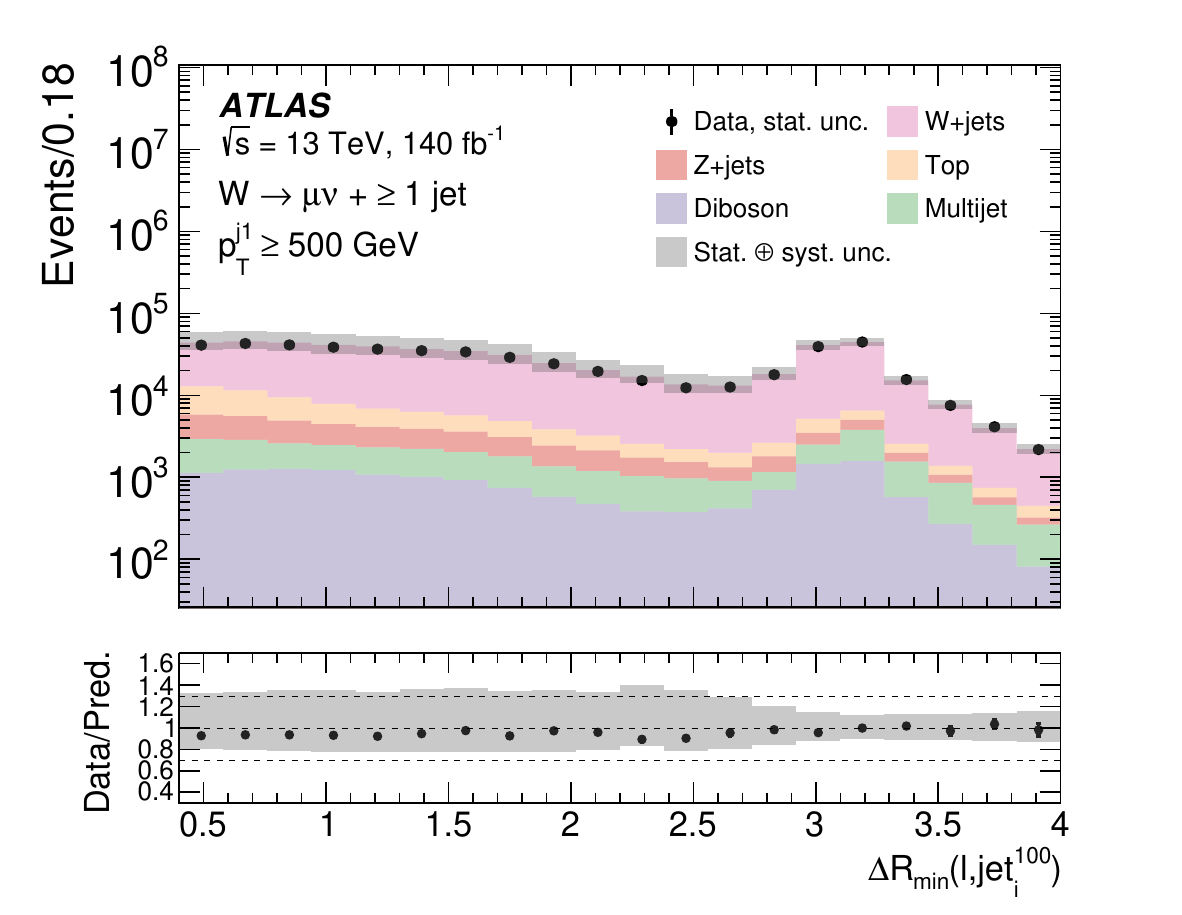}} \\

\subfloat[]{\includegraphics[width=0.45\textwidth]{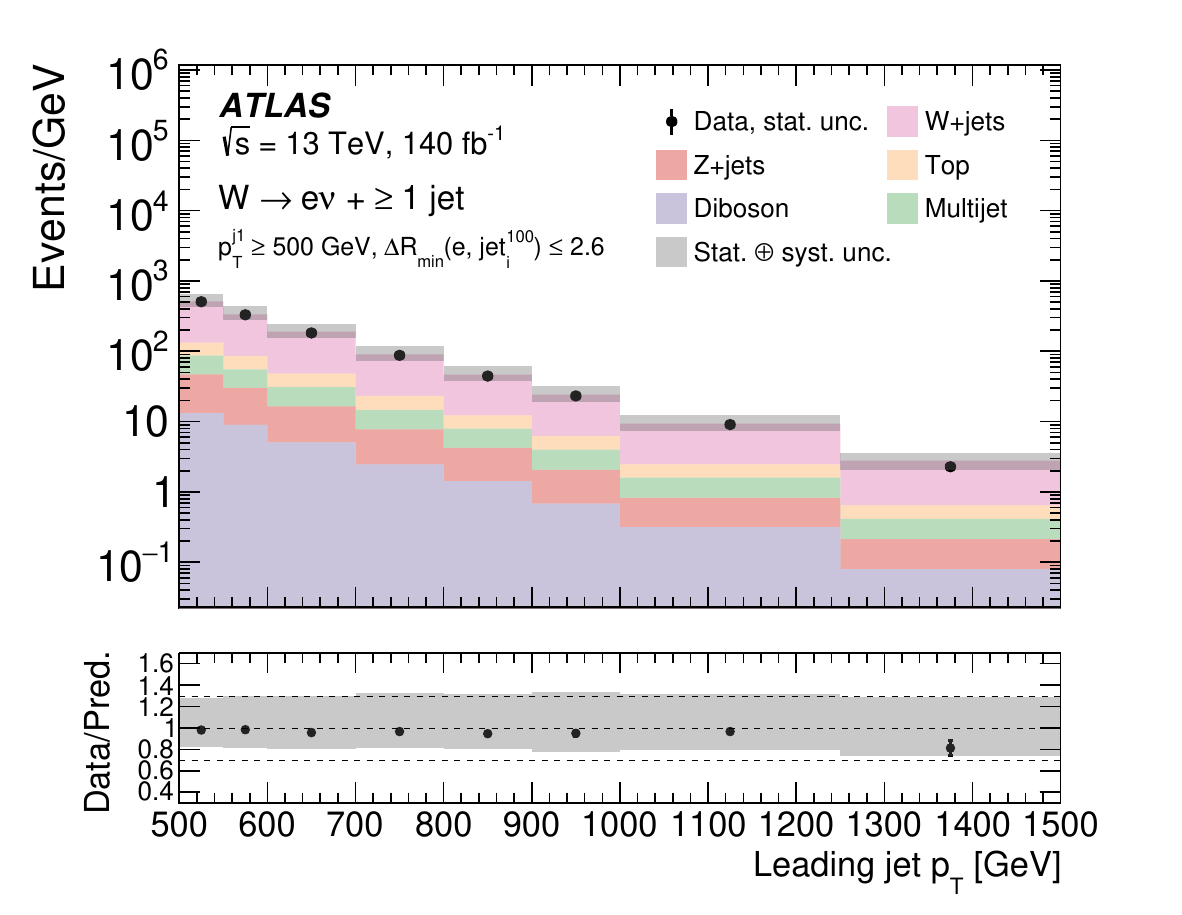}}\qquad
\subfloat[]{\includegraphics[width=0.45\textwidth]{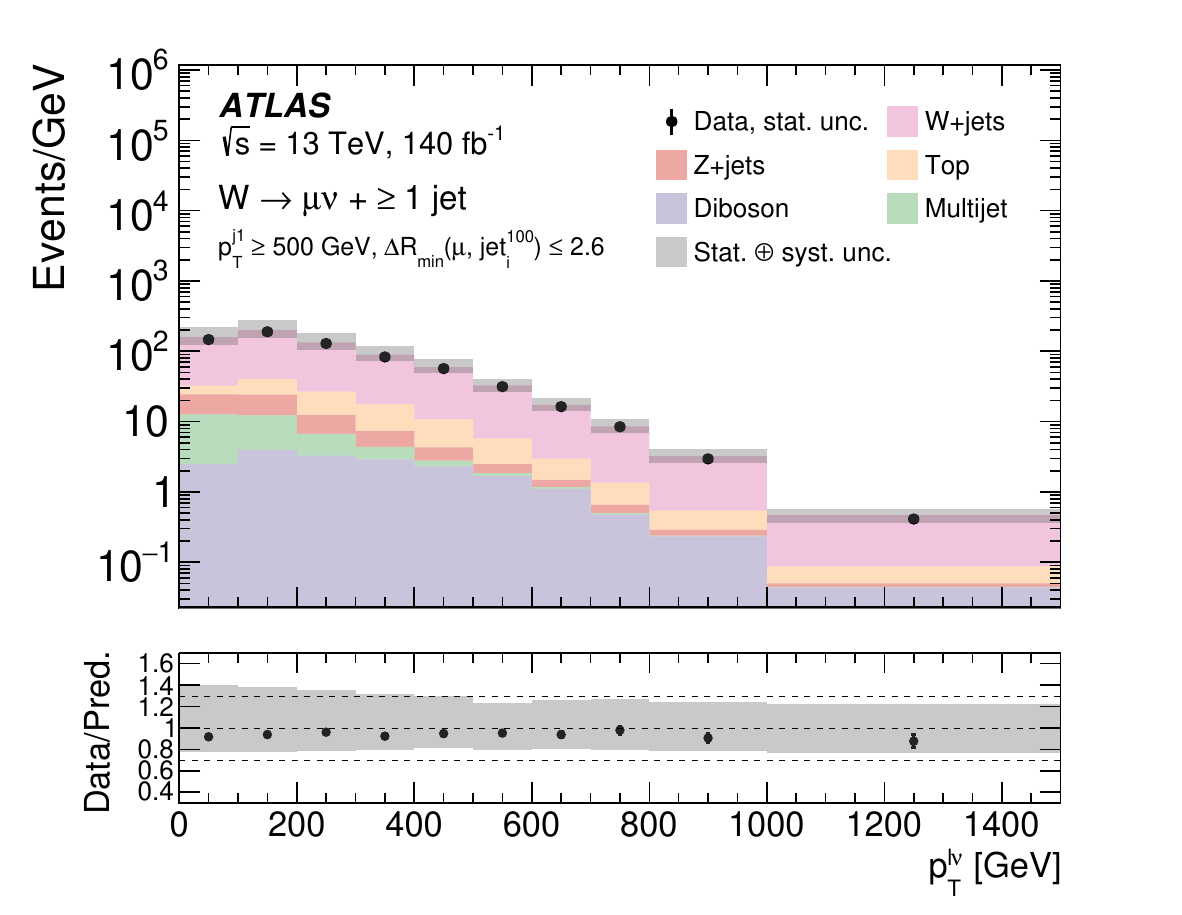}}
\caption{
Differential distributions at reconstruction level in the (a, c) electron or (b, d) muon channel for (a, b) inclusive and (c, d) collinear signal regions
after the application of the background normalisation factors.
The signal process is stacked above all background predictions.
The bottom panel shows the ratio of the data to the total signal plus background prediction.
The shaded band includes statistical and systematic uncertainties from signal and background processes added in quadrature.
}
\label{fig:backgrounds:reco}
\end{center}
\end{figure}


\section{Unfolding}
\label{sec:unfolding}

The \xsec measurements unfolded to particle level using an iterative Bayesian unfolding technique~\cite{DAgostini:1995}
to account for object and selection inefficiencies, small acceptance corrections, and resolution effects.
The data are unfolded to particle-level regions that closely match the reconstruction-level object and phase-space selections shown in Table~\ref{selection:recoSelection}.
Final-state photons radiated from the leptons are added to the lepton four-momentum within a cone of $\Delta R = 0.1$.
These leptons are referred to as `dressed leptons'.
The \antikt jet clustering algorithm is used to cluster all final state particles with a lifetime greater than 30~ps into jets,
excluding the neutrino and the electron or muon from the \Wboson decay and any photon included in the dressed lepton.
The missing transverse momentum is computed from the final-state neutrino from the \Wboson decay.
Events are then selected to contain at least one particle-level jet with transverse momentum greater than 500~\GeV
and exactly one dressed lepton with transverse momentum greater than 30~\GeV and $|\eta| < 2.5$.
The lepton $\eta$ requirements are harmonised between the electron and muon channels to unfold to a common fiducial region.
All other requirements are applied that define the regions in Table~\ref{selection:recoSelection},
except for the \bjet veto requirement to keep the particle-level measurement inclusive in jet flavour.
The unfolding algorithm implemented in the \textsc{RooUnfold} toolkit~\cite{Adye:2011gm} is used with two iterations.
The electron and muon channels are unfolded separately and then combined as discussed in Section~\ref{sec:results}.

The signal event yields are determined by subtracting the estimated background contributions from the data satisfying the detector-level selection.
The resulting distributions are corrected for detector-level effects to the fiducial phase space at particle level.
The \SHERPA[2.2.11]  generator is used as the nominal signal prediction in the unfolding procedure.
Simulated signal events, satisfying both the detector-level and particle-level selections,
are used to generate a response matrix for each distribution and the particle-level distribution
is used as the initial prior to determine the first estimate of the unfolded data distribution.
For the second iteration, the new estimate of unfolded data is obtained using the background-subtracted data and an unfolding matrix,
which is derived using Bayes’ theorem, from the response matrix and the current prior.
Before entering the iterative unfolding, the background-subtracted data are corrected for the expected fraction of events
that satisfy the detector-level selection, but not the particle-level one (unmatched-events).
For each bin of each differential distribution, the unfolded event yields are divided by the
integrated luminosity of the data sample and by the bin width, to obtain the measured \xsec.

A $\Delta R$-matching procedure is applied to fill the matrices and the unmatched-events fraction is defined accordingly.
The lepton and highest \pt jet at reconstruction and generator level are required to be matched within an angular cone of $\Delta R=0.4$.
Moreover the reconstructed jet that is found to be closest to the lepton is required to be matched to the closest particle-level jet within an angular cone of $\Delta R=0.4$.
In the case of the inclusive-2 jets region, the sub-leading jet at reconstruction level is also required to be matched to the sub-leading jet at particle level.
If the matching criteria fail, the corresponding event does not enter the response matrix calculation and is considered
as an unmatched fake object that is subtracted from the data alongside other backgrounds before unfolding.

Jets that satisfy the generator-level selection on the leading jet of \pt{}$>$~500~\GeV
may fail this requirement at reconstruction level due to the a finite jet energy resolution.
To account for these migration effects, an additional underflow bin is added for all observables,
where the selection on the reconstruction level leading jet transverse momentum is relaxed to $400<\pt{}<500$~\GeV.
These migrations account for less than 10\% of the total events in the fiducial region.



\section{Systematic uncertainties}
\label{sec:syst}

Several sources of experimental and theoretical uncertainties are considered during the unfolding process.
Systematic uncertainties are derived for each observable by propagating changes from each
systematic source through the MC unfolding inputs and the subtracted background in the unfolded data.
For all sources of uncertainty that affect the background model, the background estimate is recalculated
with the modified background model and propagated to the subtracted background before unfolding.
For experimental uncertainties affecting the reconstructed signal and background in a similar way,
the unfolding response matrices, reconstruction efficiencies and the unmatched events corrections
are adjusted alongside the subtracted background in the unfolded data.
Uncertainties are assessed separately for the electron and muon channels,
with correlations taken into account as necessary.

Systematic uncertainties relating to the lepton triggering, reconstruction, identification, and isolation are included~\cite{EGAM-2018-01,EGAM-2021-01,MUON-2018-03}.
Muon reconstruction uncertainties include variations in muon momentum scale, the ID and MS resolution, and the sagitta-bias corrections~\cite{MUON-2022-01}.
Variations of the electron energy scale and resolution are included and arise from
material interactions, in situ calibrations, shower shapes, and pile-up modelling~\cite{ATLAS:2023mnw}.
Electron and muon identification uncertainties account for variations of the efficiency scale factors
by varying the statistical and systematic components of the scale factors.

The jet energy scale (JES) and resolution (JER) uncertainties are derived as a function of the jet transverse momentum, rapidity, and flavour~\cite{JETM-2018-05}.
A reduction scheme is applied resulting in 29 nuisance parameters for the JES
and 13 nuisance parameters for the JER\@.
Aspects of the JES/JER uncertainty that rely on the relative fraction of quark- to gluon-initiated jets
are estimated as a function of the jet momentum and rapidity
from simulation.
The uncertainty in the efficiency corrections for the JVT requirement that is used to mitigate the impact of pile-up jets
is evaluated by shifting the scale factors by the corresponding uncertainties.
Uncertainties in the correction factors for the $b$-tagging efficiency are applied to the MC samples using
dedicated flavour-enriched samples in data and an additional term is included to extrapolate the measured
uncertainties to the very high-\pt{} phase space using MC-based uncertainties~\cite{FTAG-2018-01,FTAG-2019-02,FTAG-2020-08}.

Variations in the energy scale and resolution of all physics objects are
also propagated to the reconstruction of the missing transverse momentum.

Sources of uncertainty relating to the background estimation include
statistical uncertainties in the background normalisation factors,
theoretical uncertainties including variations of the renormalisation
and factorisation scales and PDF eigenvector variations for each background.
Additional uncertainties are applied to the \mjj{} background model for \ttbar{} and \Zjets processes.
Residual mis-modelling of the \mjj{} distribution in the control regions persists at the 1--2$\sigma$ level for \mjj$>1$~\TeV.
No additional uncertainties are added to the multijet process as described in~\Cref{sec:backgrounds} given the good modelling observed in the corresponding multijet validation region.
The difference between the nominal prediction and the data is symmetrised and taken as a conservative uncertainty.
Due to the small contributions of these backgrounds in the \mjj$>1$~\TeV measurement regions,
this uncertainty has a very small impact on the final results.

The uncertainty in the combined 2015--2018 integrated luminosity is 0.83\% \cite{DAPR-2021-01}.
The uncertainty in pile-up effects is assessed by varying the average number of pile-up interactions.

Several additional sources of uncertainties are assigned to the unfolding procedure,
which account for possible simulation or theoretical biases:

\begin{itemize}
\item \textit{Prior-related unfolding bias}: These assess the sensitivity of the unfolded data to the choice of the initial prior from the nominal \SHERPA[2.2.11] signal model.
The \SHERPA[2.2.11] sample is reweighted at generator level such that it matches the background-subtracted data at reconstruction level for each measured observable.
This pseudo-data sample is unfolded with the nominal response matrix, and the basic unfolding bias uncertainty is obtained from the difference between
the unfolded distribution and the reweighted generator-level distribution.
\item \textit{Uncertainty in migrations, efficiencies and fakes modelling}: The alternative \mgamc{} +\pythia sample is used to assess
the uncertainty due to the mis-modelling of the migrations, the reconstruction  efficiency and the fake corrections.
In this procedure, first the alternative signal sample is reweighted to match the nominal \SHERPA[2.2.11] sample
at generator level to avoid double-counting with the prior-related uncertainty.
Then the reweighted sample is used to unfold the data and the difference between that and the nominal unfolding result is taken as the systematic error.
\item \textit{Signal modelling uncertainty}: An uncertainty in the measured \xsec is assessed from signal theory uncertainties.
The scale and PDF variations in the nominal \SHERPA[2.2.11] prediction are varied in the unfolding MC inputs,
and the unfolded data are compared with the nominal unfolded \xsec in data to derive an uncertainty.
This uncertainty is negligible compared to the two other uncertainties.
\end{itemize}

The statistical uncertainty in the unfolded measurement and its bin-by-bin correlations are
assessed using the \textsl{bootstrap method}~\cite{ATL-PHYS-PUB-2021-011}
with an ensemble of 10,000 replicas of the original data sample for each measured observable.

Measurements in the electron and muon channels are combined and an average value is reported, as described in~\Cref{sec:results}.
During the combination, common sources of uncertainties are correlated across the channels, while those uniquely affecting the electron or muon channel are treated as uncorrelated.
The statistical and systematic uncertainties in the unfolded inclusive \xsecs after the combination are shown in Table~\ref{tab:systematics}.
Uncertainties in the inclusive \xsecs across all regions are approximately 3\%--4\% and dominated by systematic uncertainties,
while statistical uncertainties constitute less than 0.5\% of the uncertainty in the measured \xsec.
The dominant sources of systematic uncertainty arise from the background modelling driven by the statistical uncertainty in the normalisation factors, JES/JER, $b$-tagging, and unfolding uncertainties.
The `Others' category contains sub-dominant sources of uncertainties arising from
missing transverse momentum reconstruction and the jet-to-vertex fraction uncertainties.

Differential distributions of the relative statistical and systematic uncertainties in the unfolded \xsec for \minDRLepJet and \pTW in the inclusive region; \wptratio and leading jet \pt in the collinear region are shown in Figure~\ref{fig:syst:reco:measure_wjets:inc}.
The total uncertainty is up to 10\% in the inclusive region and up to 15\% in the collinear region.
The background modelling uncertainties are the dominant source of uncertainty and contribute at a similar level to the statistical uncertainty across most distributions.
In the back-to-back region and highest energy bins of the leading jet \pt{} and the \Wboson \pt{}
the statistical uncertainties are one of the leading sources of uncertainty and are around 10\%--15\%.
All other sources of uncertainty contribute at the few percent level across all distributions.

\begin{table}
\sisetup{group-minimum-digits=4}
\centering
\caption{Relative uncertainties (in \%) in the measured integrated \xsecs.
Statistical uncertainties from the data and sources of systematic uncertainties are shown separately.
The `Others' category contains sub-dominant uncertainties arising from missing transverse momentum reconstruction and the jet-to-vertex fraction uncertainties.}
\begin{tabular}{lllll}
\toprule
& \multicolumn{4}{c}{\textsl{Regions}}                                                       \\
Uncertainty source   & Inclusive                            & Inclusive 2-jet & Collinear       & Back-to-back    \\
\midrule
Background modelling & \numRF[2]{2.02}                      & \numRF[2]{2.06} & \numRF[2]{2.02} & \numRF[2]{2.28} \\
JES/JER              & \numRF[2]{1.25}                      & \numRF[2]{1.23} & \numRF[2]{1.11} & \numRF[2]{1.64} \\
\btag                & \numRF[2]{1.24}                      & \numRF[2]{1.37} & \numRF[2]{1.46} & \numRF[1]{0.71} \\
Lepton               & \numRF[1]{0.81}                      & \numRF[1]{0.77} & \numRF[1]{0.70} & \numRF[2]{1.15} \\
Unfolding            & \numRF[2]{0.34}                      & \numRF[1]{0.38} & \numRF[1]{0.56} & \numRF[1]{0.05} \\
Luminosity           & \numRF[1]{0.83}                      & \numRF[1]{0.83} & \numRF[1]{0.83} & \numRF[1]{0.83} \\
Pile-up              & \numRF[2]{0.19}                      & \numRF[2]{0.15} & \numRF[2]{0.22} & \numRF[2]{0.18} \\
Others               & \numRF[2]{0.14}                      & \numRF[2]{0.19} & \numRF[2]{0.20} & \numRF[1]{0.03} \\
\midrule
Data statistical     & \numRF[1]{0.43}                      & \numRF[1]{0.46} & \numRF[1]{0.51} & \numRF[1]{0.78} \\
\midrule
Total uncertainty    & \numRF[2]{3.30}                      & \numRF[2]{3.40} & \numRF[2]{3.35} & \numRF[1]{3.59} \\
\bottomrule
\label{tab:systematics}
\end{tabular}
\end{table}

\begin{figure}[!htbp]
\graphicspath{{./figures/systematics/}}
\begin{center}
\subfloat[]{\includegraphics[width=0.45\textwidth]{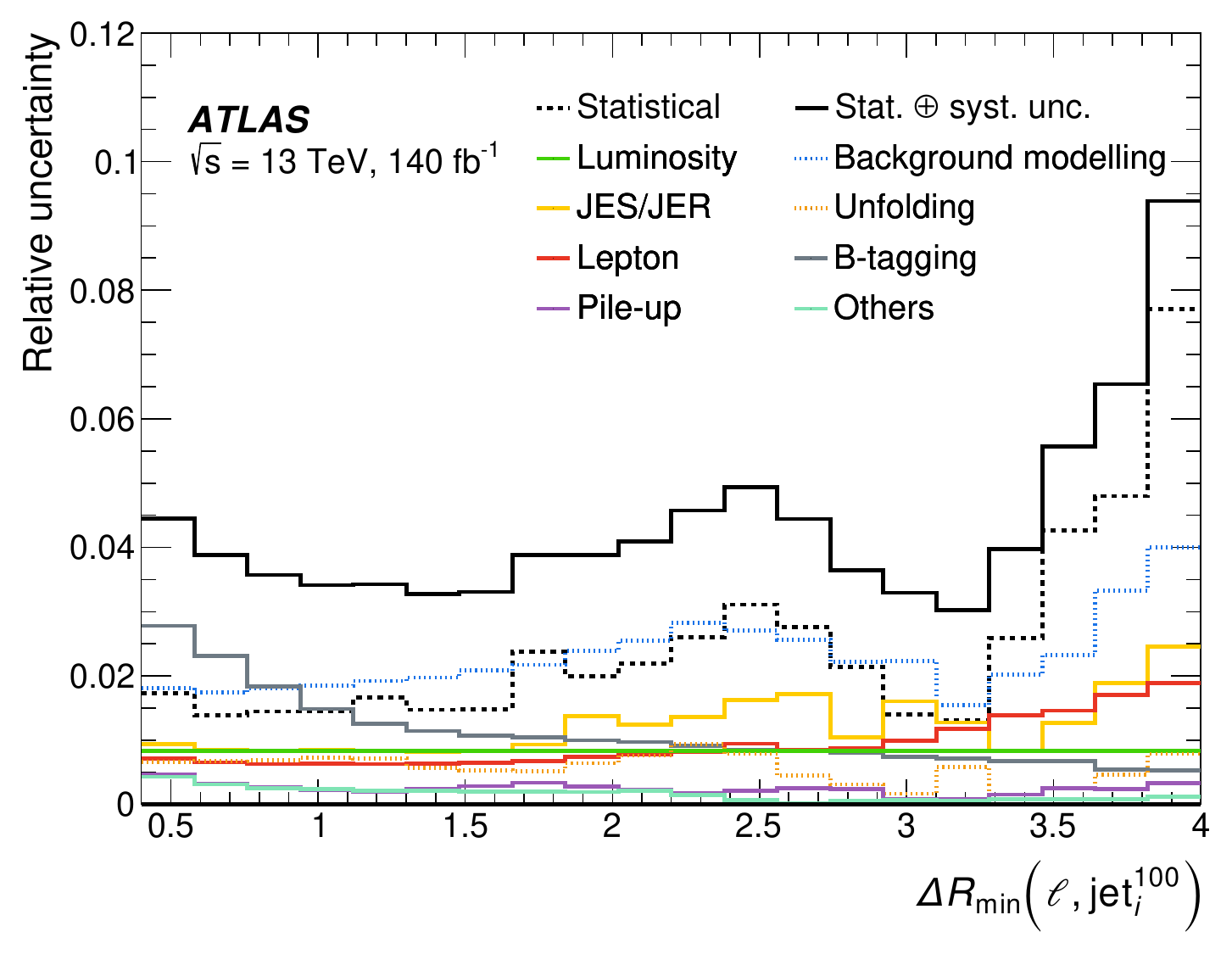}}\qquad
\subfloat[]{\includegraphics[width=0.45\textwidth]{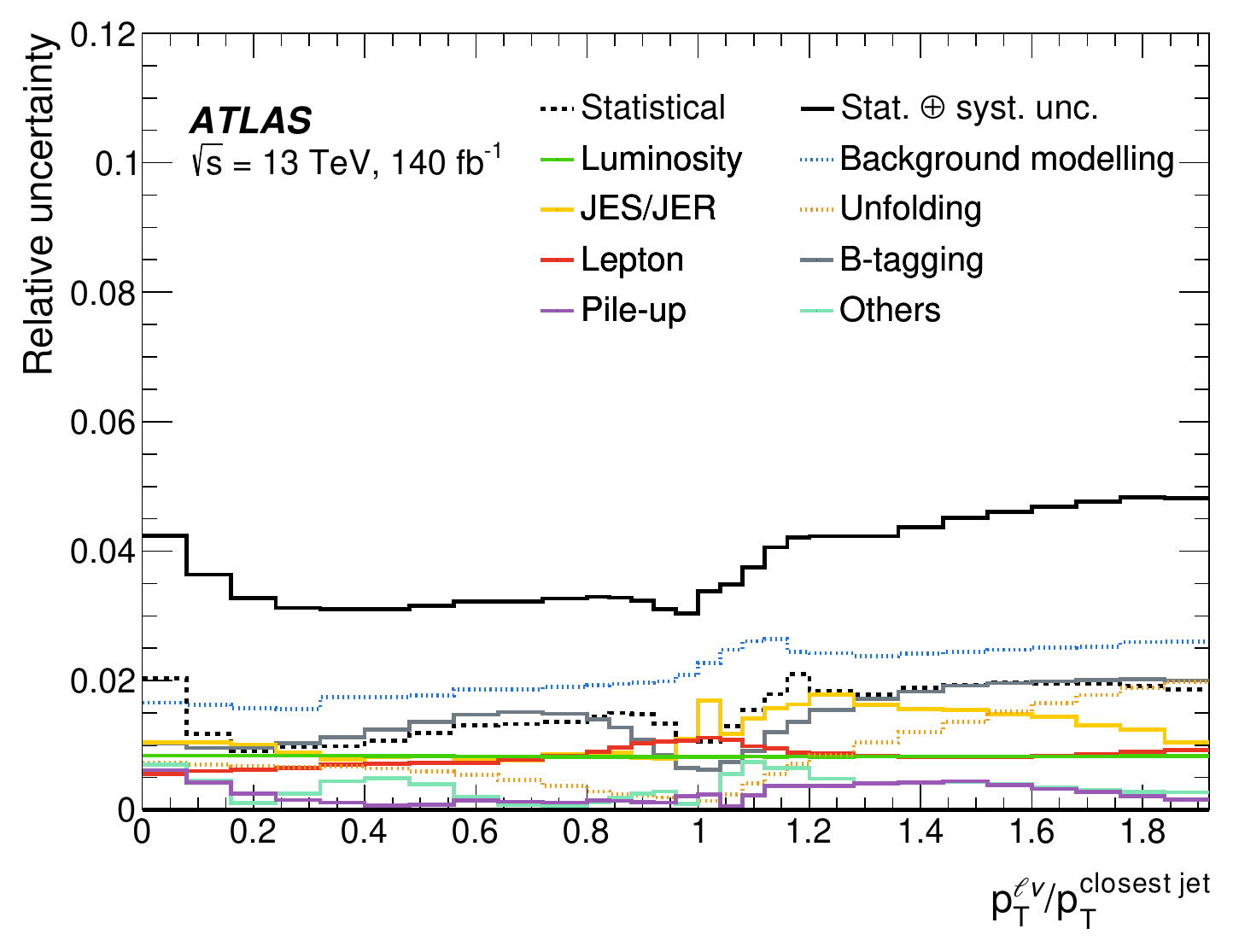}}   \\
\subfloat[]{\includegraphics[width=0.45\textwidth]{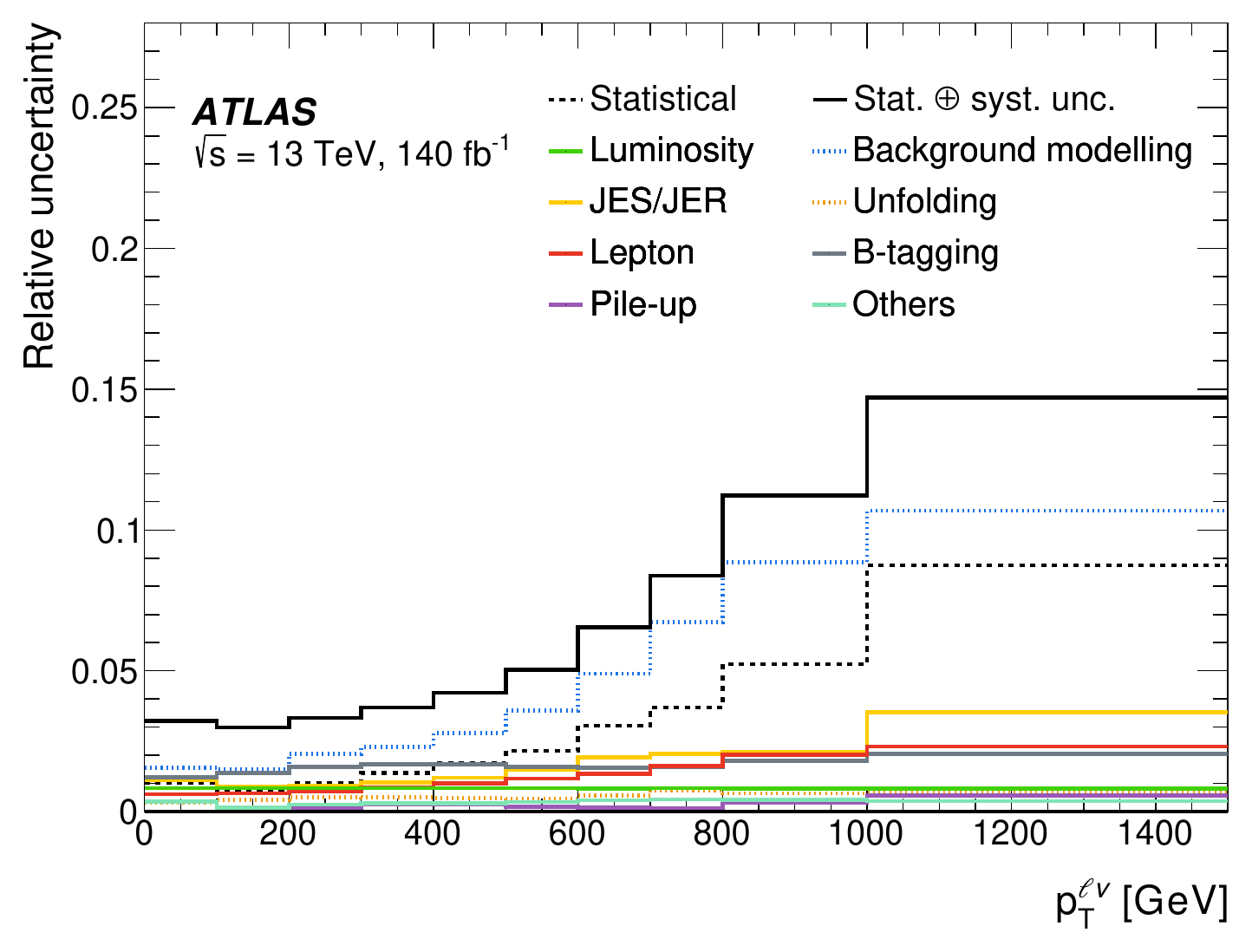}}\qquad
\subfloat[]{\includegraphics[width=0.45\textwidth]{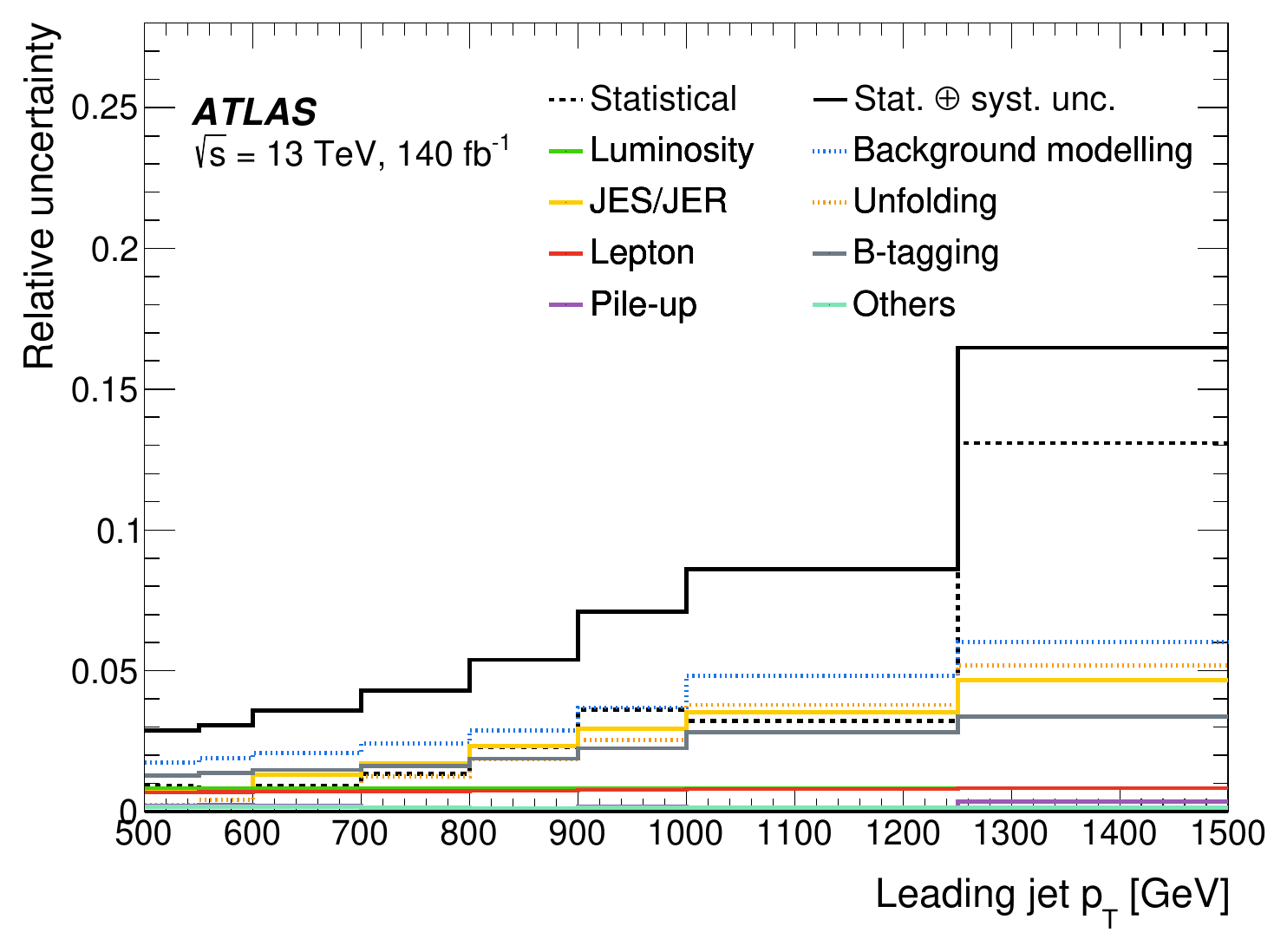}}
\caption{
Relative systematic uncertainties in the averaged \xsec
for various differential distributions in the (a, b) inclusive and (c, d) collinear phase spaces.
The upper solid line shows the total uncertainty in the measured \xsec in data,
and includes correlations between the systematic components.
The `Others' category contains sub-dominant uncertainties arising from missing transverse momentum reconstruction and the jet-to-vertex fraction uncertainties.
}
\label{fig:syst:reco:measure_wjets:inc}
\end{center}
\end{figure}



%
\section{Results}
\label{sec:results}

The \xsecs in the electron and muon channels are measured separately
and an average \xsec is obtained using a $\chi^2$ likelihood fit with the \textsc{Combiner}~\cite{dado_2024_12007778} package.
In this method, all uncertainties are treated as correlated between the electron and muon channels,
except for statistical uncertainties and lepton-specific systematic uncertainties such as unfolding uncertainties, lepton triggering, and the lepton energy and momentum resolution.
Additionally, uncertainties in the multijet backgrounds are considered uncorrelated
since the two channels are dominated by different fake-lepton processes.
All systematic uncertainties are symmetrised before the combination.
The combined \xsec is found to be consistent with the individual
electron and muon channels for the full set of regions and observables.

The measured data is compared with several state-of-the-art theoretical predictions,
including two NLO multijet merged generator predictions: the \SHERPA[2.2.11]  and \mgamc+\pythia samples described in Section~\ref{sec:samples}.
The \SHERPA[2.2.11] predictions are shown with and without NLO EW$_{\text{virt}}$ corrections to demonstrate their impact on the measured distributions.
Additionally, a fixed-order prediction of \Wplusonejet was generated at NNLO in $\alphas$ using the
MCFM program~\cite{Campbell:2023lcy,Boughezal:2015dva,Campbell:2019dru} and is compared with the data.
The NNLO calculation relies on the $N$-jettiness subtraction formalism and the event shape variable $N$-jettiness $\tau$~\cite{Stewart:2010tn}.
For all results presented in this paper, a requirement of $\tau_\text{cut} = 10^{-4}$ is used and found to impact the final results
at the sub-percent level for all differential distributions measured.
A dynamic scale choice is used where $\mu = \sqrt{ m_{\ell\nu}^2 + \sum_{\text{jets}} p_{\text{T}}^2}$,
where $m_{\ell\nu}$ is the invariant mass of the lepton--neutrino system
and the second term contains the scalar sum of the transverse momentum squared of all jets with $\pt>30$~\GeV.
The \NNPDF[3.0nnlo] set of PDFs was used.
The PDF uncertainties from the MCFM calculation are expected to be similar in size to \SHERPA[2.2.11] NLO calculations and shown in~\Cref{tab:summary}.
The renormalisation and factorisation scales are varied coherently by a factor of two, avoiding asymmetric variations.
Non-perturbative effects in the high-momentum phase space of this analysis were found to be negligible for all measured observables.
The effects of higher-order EW corrections on the MCFM prediction are assessed by extrapolating from the \SHERPA prediction.
A bin-by-bin correction factor computed from the ratio of the \SHERPA QCD with EW$_{\text{virt}}$ prediction to the QCD-only prediction
is applied to the MCFM prediction and displayed separately throughout all results.
This procedure was discussed with the MCFM authors and also done in the corresponding \Zjets analysis~\cite{ATLAS:2022nrp}.

The combined inclusive \xsecs in the inclusive, inclusive 2-jet, collinear, and back-to-back regions
are shown in Figure~\ref{fig:unfolding:unfolded:summary} and tabulated in Table~\ref{tab:summary}.
For results showing predictions accurate to QCD, uncertainties include scale and PDF uncertainties added in quadrature,
while the results reported with EW virtual corrections display uncertainties arising from the different EW combination schemes only.
The measurement uncertainties are significantly smaller than the uncertainties in the
signal predictions from the \SHERPA[2.2.11] and \mgamc+\pythia event generators,
which are dominated by variations in the renormalisation and factorisation scale uncertainties.
In particular, \SHERPA[2.2.11] uncertainties are larger than the \mgamc{} uncertainties due to the 3-5j@LO calculations of the matrix element for \SHERPA.
The uncertainty shown in~\Cref{fig:unfolding:unfolded:summary} is the
total uncertainty in the prediction which is very large in comparison to the
EW corrections and scale uncertainties in the QCD predictions.
The fixed-order MCFM \Wplusonejet prediction computed at NNLO accuracy
is found to be in good agreement with the data across all regions
with a precision comparable to that of the measured \xsecs from data.
The NLO EW$_{\text{virt}}$ corrections lead to an overall decrease in the inclusive \xsec by 10\%--12\%,
in agreement with fixed-order calculations computed for $V$+2-jet in Ref.~\cite{Kallweit:2015dum}.
This improves the agreement of the \SHERPA prediction with the data in the collinear region,
but leads to an approximate 10\% underestimate of the \xsecs for the MCFM prediction.

\begin{figure}[!t]
\graphicspath{{./figures/results/}}
\begin{center}
\includegraphics[width=0.8\textwidth]{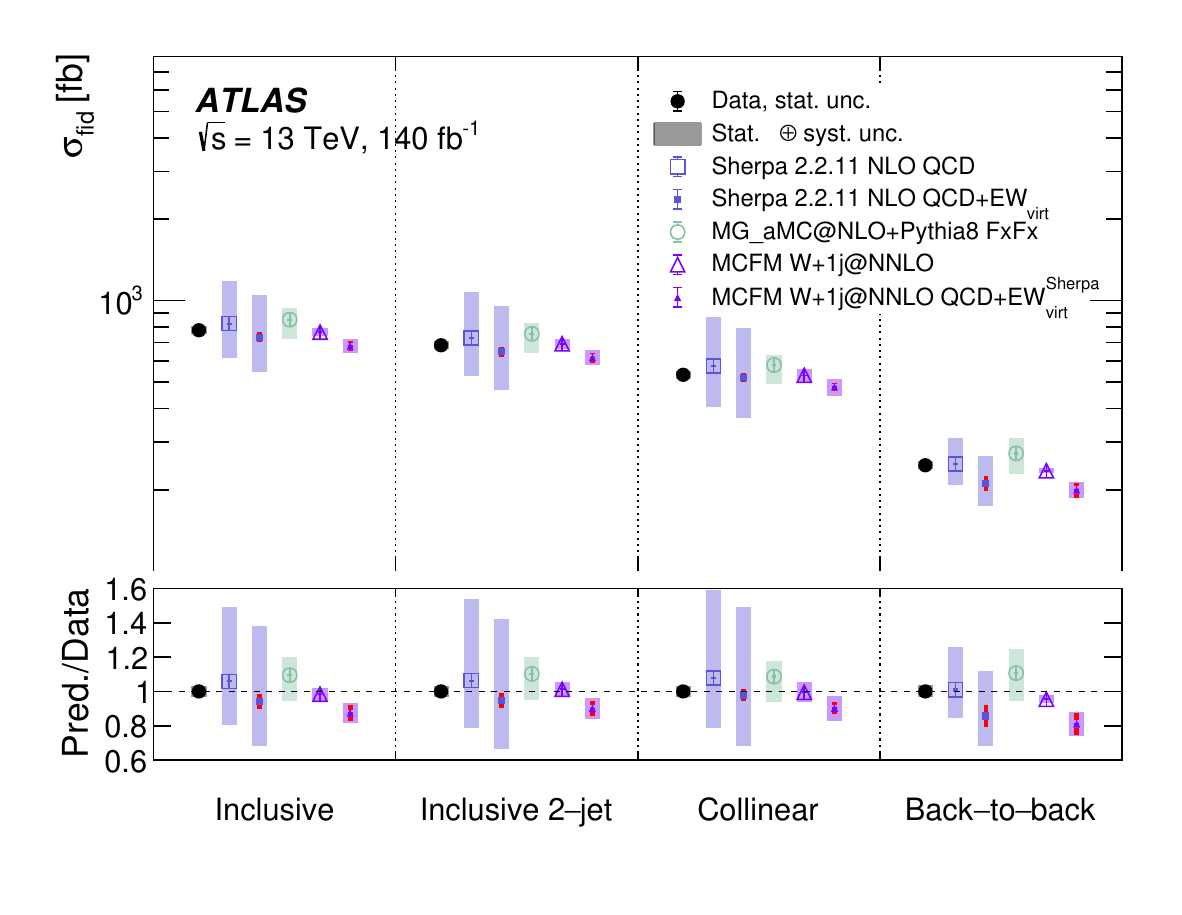} \\
\caption{
Measured fiducial \xsecs in each signal region. The measured \xsecs in data and their total uncertainty are shown by the solid dots and error band.
Various theoretical predictions are overlaid and compared with the data in the lower panel.
The small error bars inside the markers are the statistical uncertainties.
The large shaded areas are the theoretical systematic uncertainties.
Uncertainties in predictions accurate to NLO precision include QCD scale and PDF uncertainties,
while predictions that add NLO EW$_{\text{virt}}$ (electroweak) corrections show uncertainties only due to different electroweak combination schemes.
Uncertainties in predictions accurate at fixed-order NNLO precision are derived primarily from MCFM, do not include PDF variations, with an additional extrapolation from the \SHERPA QCD with EW$_{\text{virt}}$ prediction to assess the effects of higher-order electroweak corrections.
On predictions with EW$_{\text{virt}}$ corrections, the total uncertainty is shown as a shaded region,
while the electroweak correction is overlaid with an error bar line.
}
\label{fig:unfolding:unfolded:summary}
\end{center}
\end{figure}

\begin{table}[!htbp]
\sisetup{group-minimum-digits=4}
\setlength\doublerulesep{1.0pt}
\centering
\renewcommand{\arraystretch}{1.22}
\caption{
Measured fiducial \xsec ($\sigma_\text{fid}$) in each signal region with theoretical predictions from \SHERPA[2.2.11], \mgamc+\pythia, and MCFM W+1-jet@NNLO.
Systematic uncertainties in the measured \xsecs are separated into statistical and non-statistical sources,
while theoretical predictions show uncertainties arising from scale and PDF variations separately.
For results showing predictions accurate to QCD, uncertainties include scale and PDF uncertainties added in quadrature,
while the results reported with EW virtual corrections display uncertainties arising from the different EW combination schemes only.
}
\resizebox{1.0\width}{!}{
\begin{tabular}{lllll}
\toprule[1pt]
\midrule[0.5pt]
\textbf{Inclusive}       & \multicolumn{2}{c}{$\sigma_\text{fid}$ (fb)} & \multicolumn{2}{c}{Total $\Delta\sigma_\text{fid}$ (fb)}                                                                     \\
\midrule
Data                     & \numRF[3]{777.7}                             & $\pm \numRF[1]{3.3}$ (stat.)                             &                             & $\pm \numRF[2]{25.2}$ (syst.+stat.) \\
\midrule
\SHERPA[2.2.11]          & \numRF[3]{822.9}                             & $\substack{+\numRF[1]{7.7}                                                                                                   \\ -\numRF[2]{10.4}}$ (PDF) & $\substack{+\numRF[2]{357.0} \\ -\numRF[2]{208.4}}$ (scale) & $\substack{+\numRF[2]{357.6} \\ -\numRF[2]{209.4}}$ \\
\SHERPA[2.2.11] QCD+EW   & \numRF[3]{732.5}                             &                                                          & $\substack{+\numRF[2]{31.9}                                       \\ -\numRF[2]{31.9}}$ (EW virt) & $\substack{+\numRF[2]{359.0} \\ -\numRF[2]{211.8}}$ \\
\mgamc+\pythia           & \numRF[3]{851.2}                             & $\substack{+\numRF[1]{5.7}                                                                                                   \\ -\numRF[1]{5.8}}$ (PDF) & $\substack{+\numRF[2]{90.0} \\ -\numRF[2]{129.3}}$ (scale) & $\substack{+\numRF[2]{90.3}  \\ -\numRF[2]{129.5}}$ \\
MCFM                     & \numRF[3]{765.4}                             &                                                          & $\substack{+\numRF[2]{25.9}                                       \\ -\numRF[2]{25.9}}$ (scale) & \ \\
MCFM QCD+EW              & \numRF[2]{679.3}                             &                                                          & $\substack{+\numRF[1]{30.3}                                       \\ -\numRF[1]{30.3}}$ (EW virt) & $\substack{+\numRF[1]{39.9} \\ -\numRF[1]{39.9}}$ \\
\midrule[0.5pt]
\midrule[0.5pt]
\textbf{Inclusive 2-jet} & \multicolumn{2}{c}{$\sigma_\text{fid}$ (fb)} & \multicolumn{2}{c}{Total $\Delta\sigma_\text{fid}$ (fb)}                                                                     \\
\midrule
Data                     & \numRF[3]{684.3}                             & $\pm \numRF[1]{3.1}$ (stat.)                             &                             & $\pm \numRF[2]{22.8}$ (syst.+stat.) \\
\midrule
\SHERPA[2.2.11]          & \numRF[3]{727.6}                             & $\substack{+\numRF[1]{7.1}                                                                                                   \\ -\numRF[1]{9.8}}$ (PDF) & $\substack{+\numRF[2]{344.7} \\ -\numRF[2]{200.8}}$ (scale) & $\substack{+\numRF[2]{345.3} \\ -\numRF[2]{201.7}}$ \\
\SHERPA[2.2.11] QCD+EW   & \numRF[3]{648.8}                             &                                                          & $\substack{+\numRF[2]{27.7}                                       \\ -\numRF[2]{27.7}}$ (EW virt) & $\substack{+\numRF[2]{346.4} \\ -\numRF[2]{203.6}}$ \\
\mgamc+\pythia           & \numRF[3]{754.2}                             & $\substack{+\numRF[1]{5.1}                                                                                                   \\ -\numRF[1]{5.1}}$ (PDF) & $\substack{+\numRF[2]{75.9} \\ -\numRF[2]{114.7}}$ (scale) & $\substack{+\numRF[2]{76.1}  \\ -\numRF[2]{114.9}}$ \\
MCFM                     & \numRF[3]{693.4}                             &                                                          & $\substack{+\numRF[2]{28.1}                                       \\ -\numRF[2]{28.1}}$ (scale) & \ \\
MCFM QCD+EW              & \numRF[2]{616.3}                             &                                                          & $\substack{+\numRF[1]{27.1}                                       \\ -\numRF[1]{27.1}}$ (EW virt) & $\substack{+\numRF[1]{39.0} \\ -\numRF[1]{39.0}}$ \\
\midrule[0.5pt]
\midrule[0.5pt]
\textbf{Collinear}       & \multicolumn{2}{c}{$\sigma_\text{fid}$ (fb)} & \multicolumn{2}{c}{Total $\Delta\sigma_\text{fid}$ (fb)}                                                                     \\
\midrule
Data                     & \numRF[3]{532.4}                             & $\pm \numRF[1]{2.7}$ (stat.)                             &                             & $\pm \numRF[2]{17.6}$ (syst.+stat.) \\
\midrule
\SHERPA[2.2.11]          & \numRF[3]{574.0}                             & $\substack{+\numRF[1]{5.5}                                                                                                   \\ -\numRF[1]{7.9}}$ (PDF) & $\substack{+\numRF[2]{294.7} \\ -\numRF[2]{167.4}}$ (scale) & $\substack{+\numRF[2]{295.1} \\ -\numRF[2]{168.2}}$ \\
\SHERPA[2.2.11] QCD+EW   & \numRF[3]{521.3}                             &                                                          & $\substack{+\numRF[2]{18.3}                                       \\ -\numRF[2]{18.3}}$ (EW virt) & $\substack{+\numRF[2]{295.7} \\ -\numRF[2]{169.2}}$ \\
\mgamc+\pythia           & \numRF[3]{578.4}                             & $\substack{+\numRF[1]{3.9}                                                                                                   \\ -\numRF[1]{3.9}}$ (PDF) & $\substack{+\numRF[1]{52.1} \\ -\numRF[1]{85.0}}$ (scale) & $\substack{+\numRF[1]{52.4}  \\ -\numRF[1]{85.1}}$ \\
MCFM                     & \numRF[3]{529.9}                             &                                                          & $\substack{+\numRF[2]{30.5}                                       \\ -\numRF[2]{30.5}}$ (scale) & \\
MCFM QCD+EW              & \numRF[3]{480.3}                             &                                                          & $\substack{+\numRF[2]{17.2}                                       \\ -\numRF[2]{17.2}}$ (EW virt) & $\substack{+\numRF[2]{35.0} \\ -\numRF[2]{35.0}}$ \\
\midrule[0.5pt]
\midrule[0.5pt]
\textbf{Back-to-back}    & \multicolumn{2}{c}{$\sigma_\text{fid}$ (fb)} & \multicolumn{2}{c}{Total $\Delta\sigma_\text{fid}$ (fb)}                                                                     \\
\midrule
Data                     & \numRF[3]{246.5}                             & $\pm \numRF[1]{1.9}$ (stat.)                             &                             & $\pm \numRF[1]{8.6}$ (syst.+stat.)  \\
\midrule
\SHERPA[2.2.11]          & \numRF[3]{248.9}                             & $\substack{+\numRF[1]{2.3}                                                                                                   \\ -\numRF[1]{2.6}}$ (PDF)      & $\substack{+\numRF[1]{62.4} \\ -\numRF[1]{41.0}}$ (scale) & $\substack{+\numRF[1]{62.5} \\ -\numRF[1]{41.2}}$ \\
\SHERPA[2.2.11] QCD+EW   & \numRF[2]{211.3}                             &                                                          & $\substack{+\numRF[1]{13.5}                                       \\ -\numRF[1]{13.5}}$ (EW virt) & $\substack{+\numRF[1]{63.9} \\ -\numRF[1]{43.4}}$ \\
\mgamc+\pythia           & \numRF[3]{272.8}                             & $\substack{+\numRF[1]{2.0}                                                                                                   \\ -\numRF[1]{2.0}}$ (PDF)      & $\substack{+\numRF[1]{37.8} \\ -\numRF[1]{44.3}}$ (scale) & $\substack{+\numRF[1]{37.9}  \\ -\numRF[1]{44.4}}$ \\
MCFM                     & \numRF[3]{235.4}                             &                                                          & $\substack{+\numRF[1]{5.3}                                        \\ -\numRF[1]{5.3}}$ (scale)    & \\
MCFM QCD+EW              & \numRF[3]{199.7}                             &                                                          & $\substack{+\numRF[2]{12.8}                                       \\ -\numRF[2]{12.8}}$ (EW virt) & $\substack{+\numRF[2]{13.9} \\ -\numRF[2]{13.9}}$ \\
\midrule[0.5pt]
\bottomrule[1pt]
\end{tabular}

}
\label{tab:summary}
\end{table}

Measurements of the differential \xsec for all observables are shown
in~\Cref{fig:unfolding:unfolded:differential:inclusive:DeltaRTruthLepJetClosest100,fig:unfolding:unfolded:differential:inclusive_2j:mjjTruth,fig:unfolding:unfolded:differential:collinear}.
The measured differential distributions are compared with the same three signal predictions as described above.
For all figures, the main panel shows the various theoretical predictions that do not include NLO EW corrections,
while the two lower panels additionally show the effect of NLO EW$_{\text{virt}}$ as computed from \SHERPA[2.2.11].
The uncertainties in the QCD accurate predictions contain uncertainties due to scale and PDF uncertainties added in quadrature,
while the predictions including NLO EW$_{\text{virt}}$ corrections display only the uncertainty due to different EW correction combination schemes.
The measured \xsecs in data and their statistical uncertainty are shown by the solid dots and error bars,
while the shaded band shows the systematic and statistical uncertainties added in quadrature.

Figure~\ref{fig:unfolding:unfolded:differential:inclusive:DeltaRTruthLepJetClosest100} shows the differential \xsecs as a function of \minDRLepJet and \wptratio in the inclusive phase space.
These observables can be used to distinguish \Wboson emission from a high-momentum jet, referred to as the collinear events,
from events where the \Wboson recoils against a single initial-state-radiation jet, referred to as back-to-back events.
Due to the leading jet $\pt>500$~\GeV requirement, these back-to-back topologies typically lead to
high-momentum \Wboson events containing a lepton whose angular separation with the leading jet is $\DRLepJetLead \sim \pi$.
Furthermore, these single-jet events contain a \Wboson with momentum that balances the leading jet in the transverse plane, leading to \wptratio close to one.
In contrast, events where the \Wboson is emitted from a high-momentum jet receive an overall collinear enhancement
in the distribution of the angular distance between the \Wboson and the closest jet.
Moreover, the largest contribution occurs when the \Wboson momentum is relatively soft and leads to \wptratio close to zero.
For larger \wptratio where the \Wboson is not quite as soft, this disagreement with the MCFM prediction is more noticeable in light of the smaller uncertainties.
These features can be clearly seen in~\Cref{fig:unfolding:unfolded:differential:inclusive:DeltaRTruthLepJetClosest100}.
The multijet merged \SHERPA and \mgamc+\pythia event generators provide an excellent description of the data across both variables.
The inclusion of NLO EW corrections in the \SHERPA prediction generally improve the agreement with the data in the collinear regions,
but lead to an underestimate of the \xsecs in the back-to-back region while remaining consistent within uncertainties of the \SHERPA prediction.
The fixed-order MCFM \Wplusonejet NNLO prediction provides a good description of the data,
except for back-to-back regions with $\minDRLepJet > \pi$. In these regions, the MCFM
prediction undershoots the data well beyond measurement and prediction uncertainties.

\begin{figure}[!tb]
\graphicspath{{./figures/results/unfold_realthang_average/}}
\begin{center}
\subfloat[]{\includegraphics[width=0.45\textwidth]{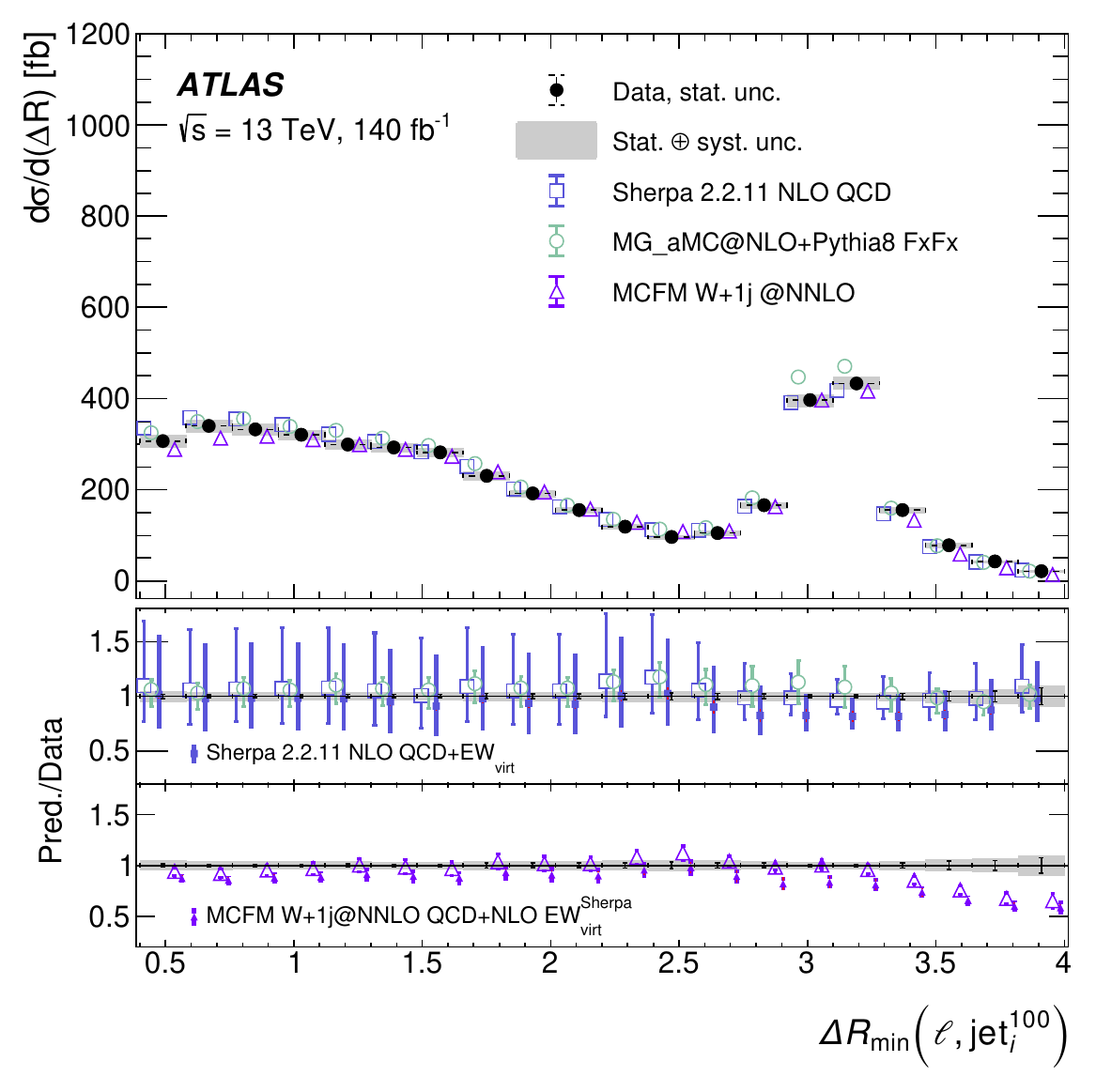}}\qquad
\subfloat[]{\includegraphics[width=0.45\textwidth]{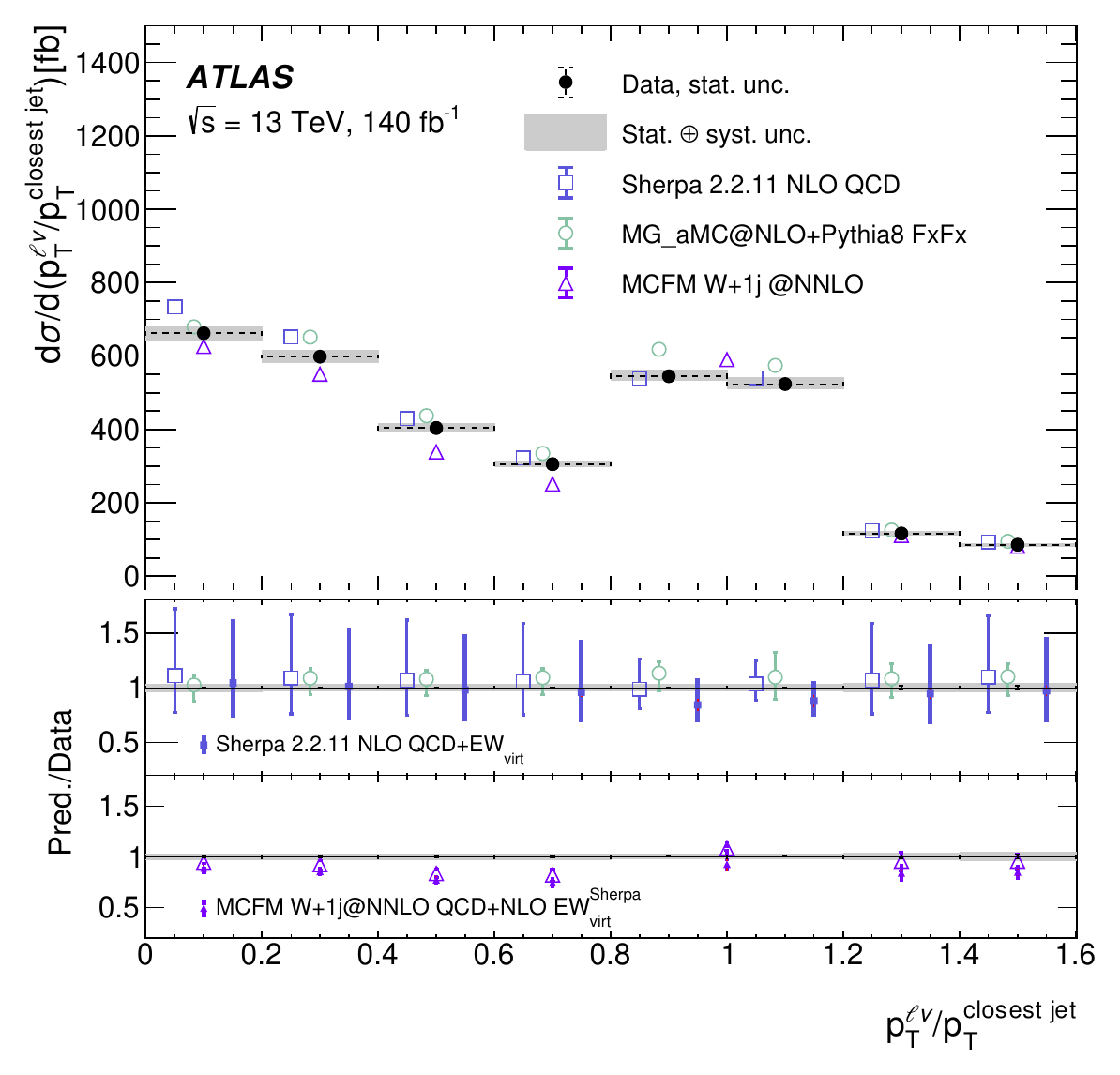}}
\caption{
Differential \xsec measurement in the inclusive phase space as function of (a) the
minimum angular separation between the lepton and any jet with transverse momentum greater than 100~\GeV ($\minDRLepJet$), and
(b) the ratio of \Wboson~\pt{} to the closest-jet \pt ($\wptratio$).
The measured \xsecs in data and their statistical uncertainty are shown by the solid dots and error bars,
while the shaded band shows the systematic and statistical uncertainties added in quadrature.
Errors bars on the theory prediction include theoretical uncertainties as discussed in the text.
The bins around \wptratio equal to 1 are merged to be insensitive to the singularity that exists
in the fixed-order MCFM calculation.
Additionally, in (b), the central two bins from the MCFM prediction are merged due to a singularity in fixed-order calculations, which requires \pt{} resummation of the \Wjets system.
The two lower panels show the effect of NLO EW$_{\text{virt}}$ as computed from \SHERPA[2.2.11].
}
\label{fig:unfolding:unfolded:differential:inclusive:DeltaRTruthLepJetClosest100}
\end{center}
\end{figure}

\Cref{fig:unfolding:unfolded:differential:inclusive_2j:mjjTruth} shows the differential \xsec
as a function of the invariant mass of the leading two jets (\mjj) in the inclusive 2-jet region.
As discussed above, the \xsec due to a relatively soft \Wboson emitted from an outgoing quark
contributes significantly to the overall production rate in the inclusive phase space.
Due to the leading jet $\pt>500$~\GeV requirement, this leads to an overall enhancement in the rate for \mjj around 1~\TeV
since the \Wboson carries only a small amount of the momentum.
Events with large \mjj{} are dominated by extremely high momentum jets with large opening angles.
This variable is an important observable for a range of BSM searches and EW-induced
measurements that have historically been difficult to model~\cite{ATLAS:2020nzk,ATLAS:2015gvj,ATLAS:2015eiz,ATLAS:2022yvh}.
Since the MCFM prediction is computed at NNLO precision for the process \Wplusonejet,
it provides NLO precision for the \mjj{} variable due to the two-jet requirement.
All theoretical predictions overestimate the \xsec around $\mjj>2$~\TeV
beyond their corresponding systematic uncertainties.

\begin{figure}[!t]
\graphicspath{{./figures/results/unfold_realthang_average/}}
\begin{center}
\includegraphics[width=0.45\textwidth]{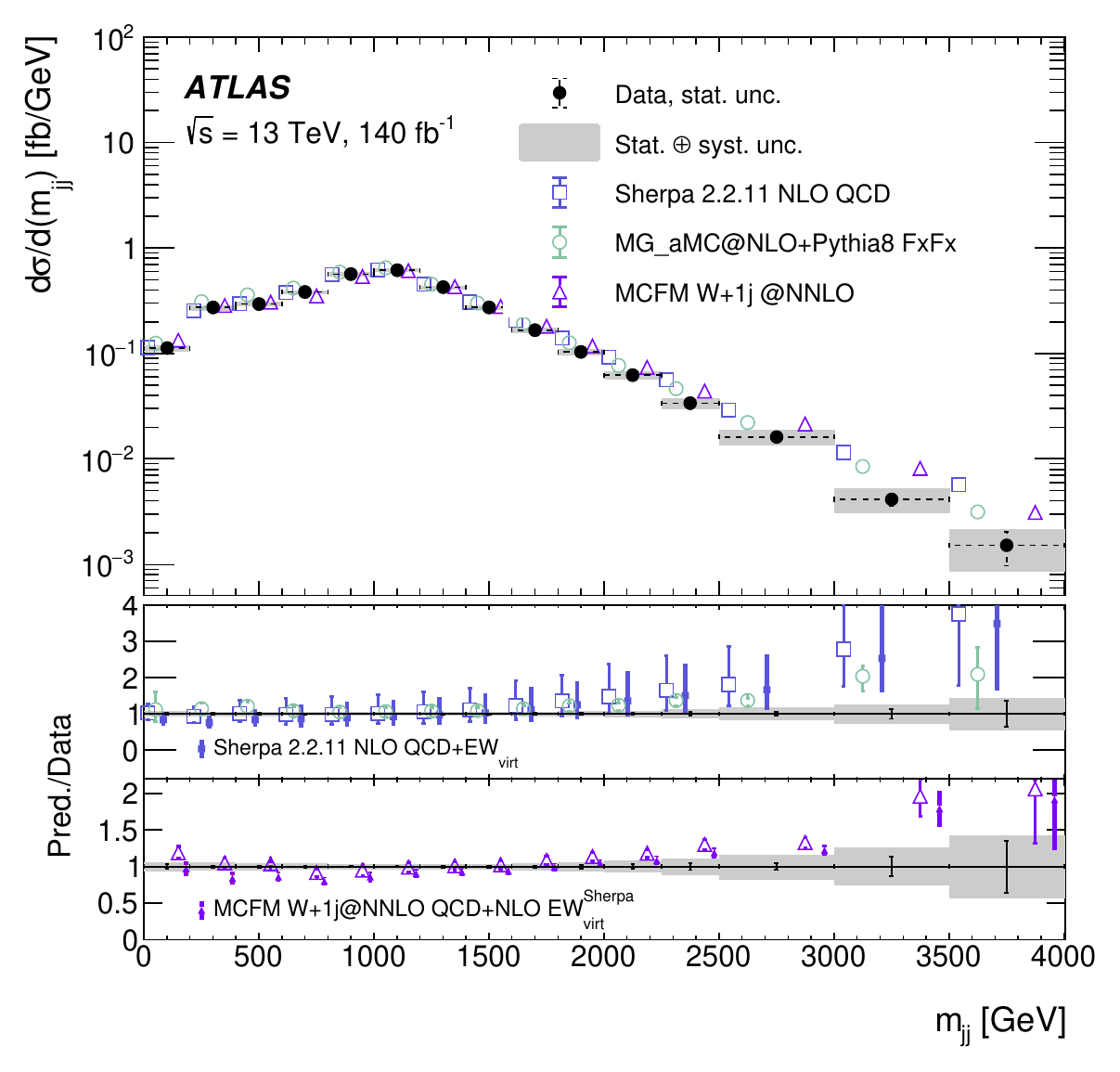}
\caption{
Differential \xsec as a function of the invariant mass of the leading two jets ($\mjj$) in the inclusive, 2-jet phase space.
The measured \xsecs in data and their statistical uncertainty are shown by the solid dots and error bars,
while the shaded band shows the systematic and statistical uncertainties added in quadrature.
Errors bars on the theory prediction include theoretical uncertainties as discussed in the text.
The two lower panels show the effect of NLO EW$_{\text{virt}}$ as computed from \SHERPA[2.2.11].
}
\label{fig:unfolding:unfolded:differential:inclusive_2j:mjjTruth}
\end{center}
\end{figure}

In the collinear selection, defined by the requirement of $\minDRLepJet<2.6$, the differential \xsec as a function of
the leading-jet transverse momentum, the inclusive jet multiplicity, the scalar sum of jet momenta,
and the \Wboson candidate transverse momentum are measured.
These measurements are shown in Figure~\ref{fig:unfolding:unfolded:differential:collinear}.

The leading jet \pt{} distribution is shown in Figure~\ref{fig:result:jet1Pt} and probes jet momenta up to the 1.5~\TeV scale.
The \mgamc+\pythia and \SHERPA predictions provide an excellent description of the data across the entire measured range.
There remain large uncertainties associated with the \SHERPA predictions even though the \SHERPA NLO EW$_{\text{virt}}$ corrections improve the agreement of the central value with the data.
The fixed-order MCFM provides an excellent description of the data with precision that matches or exceeds that of the measured data in the highest measured bins.
The application of the NLO EW$_{\text{virt}}$ corrections to the MCFM reduces the \xsec in all bins
and generally leads to a $\sim$10\% underestimate of the \xsec.

Measurements of the differential \xsec as a function of the inclusive jet multiplicity
and the scalar sum of all the jet momenta are shown in~\Cref{fig:result:njets,fig:result:ht}.
Similar to the leading jet \pt{} measurement, the multijet merged predictions provide
an excellent description of the data across all measured values.
For the most extreme \HT bins, the scale uncertainties in the \SHERPA prediction
exceed 50\% due to large contributions from the leading-order matrix elements in this region of phase space.
The fixed-order MCFM provides a good description of the data for $\HT<1.4$~\TeV but underestimates the \xsec for more extreme values.
For such extreme regions, the dominant contribution comes from events with three or more jets,  beyond the formal accuracy of the MCFM calculation.

The differential distribution of the measured \xsec as a function of \pTW is shown in~\Cref{fig:result:wpt}.
The QCD-only predictions agree well with the data at low \pTW values but lead to a small
overestimate in the high \pTW regime that is improved with inclusion of NLO EW corrections.
At high \pTW, the NLO EW$_{\text{virt}}$ corrections show the standard Sudakov behaviour and
lead to large negative corrections over the QCD-only \xsec at high transverse momentum.
The size of these corrections can be as large as 30\% and exceed uncertainties in the fixed-order MCFM NNLO prediction in the highest momentum jet bins.
The application of the NLO EW$_{\text{virt}}$ corrections significantly improves the agreement of the MCFM prediction with data in the highest \pTW bins.
Towards softer \Wboson emissions, the \SHERPA prediction is dominated by many jet events that are described by its LO accurate matrix elements
and leads to large scale uncertainties for smaller \pTW values.

\begin{figure}[!htbp]
\graphicspath{{./figures/results/unfold_realthang_average/}}
\begin{center}
\subfloat[]{\includegraphics[width=0.45\textwidth]{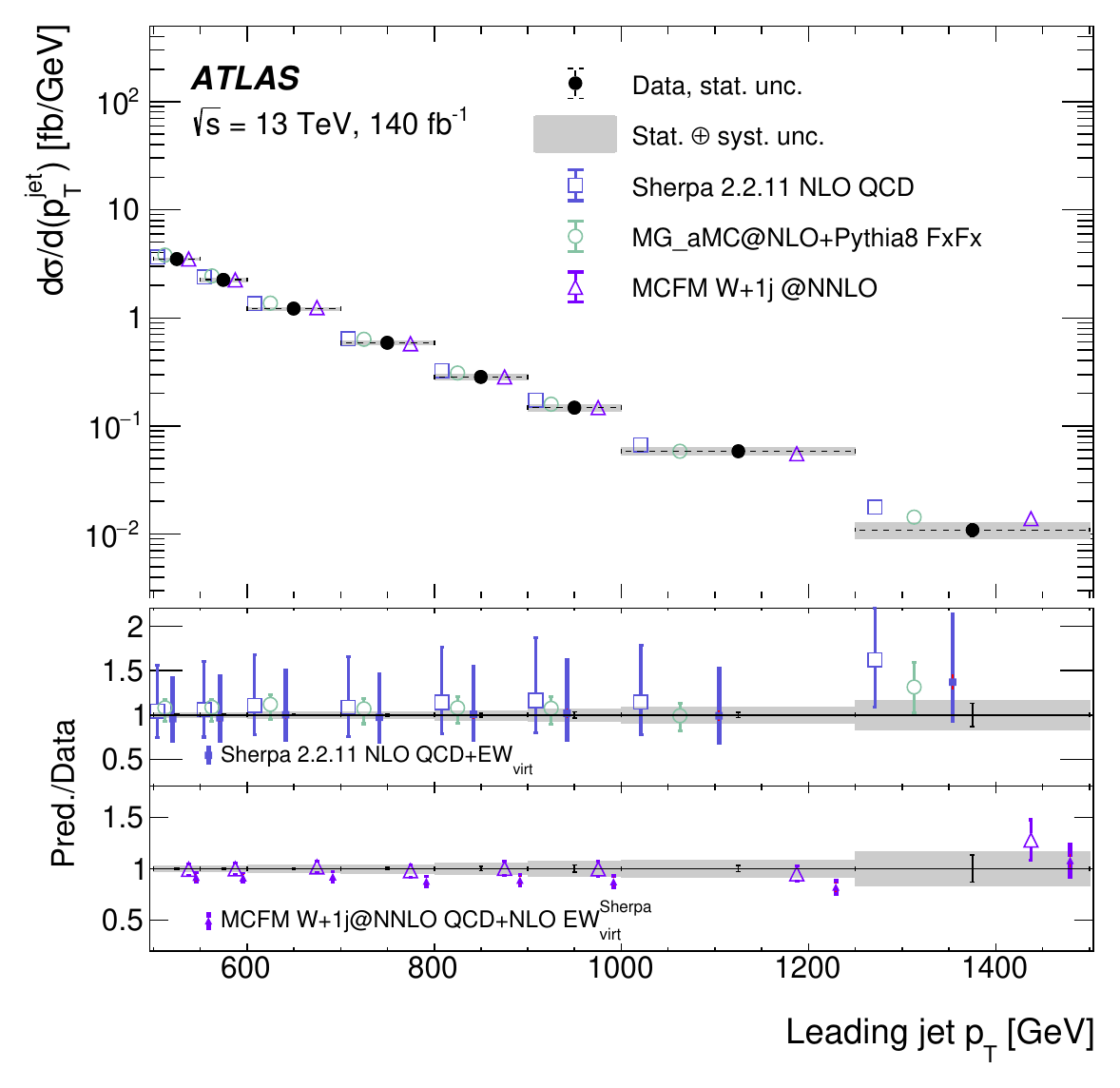} \label{fig:result:jet1Pt} }\qquad
\subfloat[]{\includegraphics[width=0.45\textwidth]{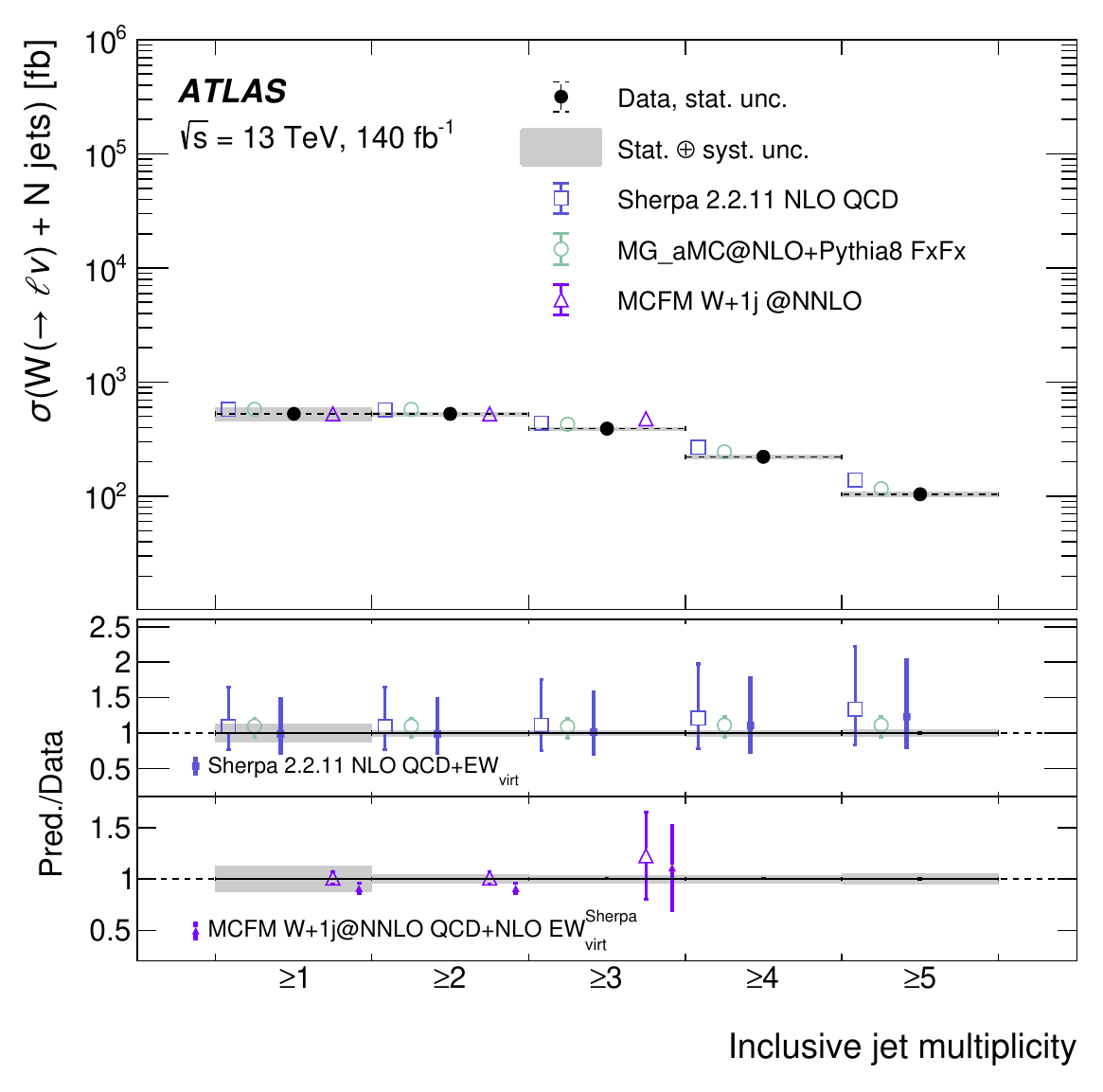} \label{fig:result:njets} }   \\
\subfloat[]{\includegraphics[width=0.45\textwidth]{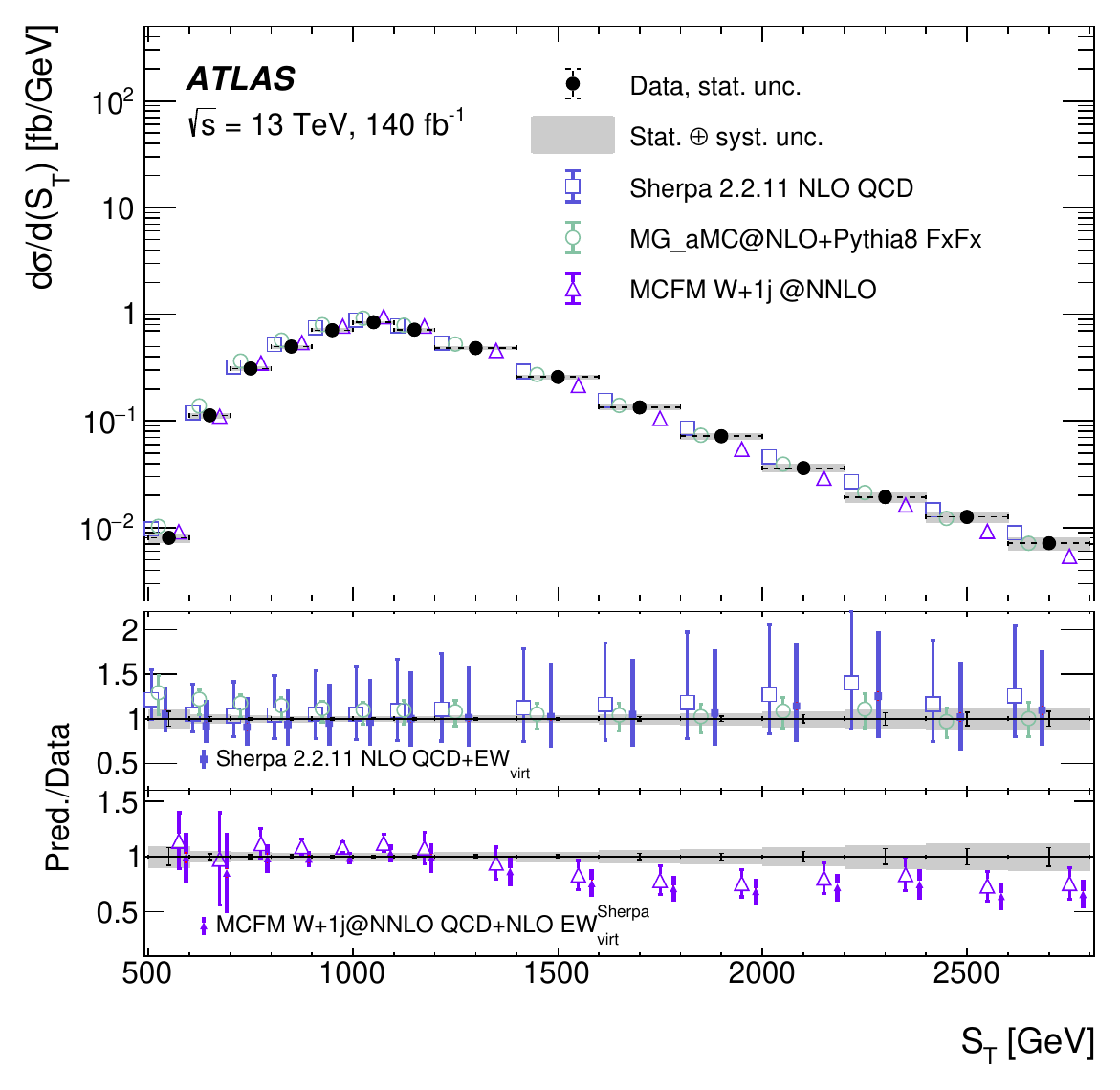} \label{fig:result:ht} }\qquad
\subfloat[]{\includegraphics[width=0.45\textwidth]{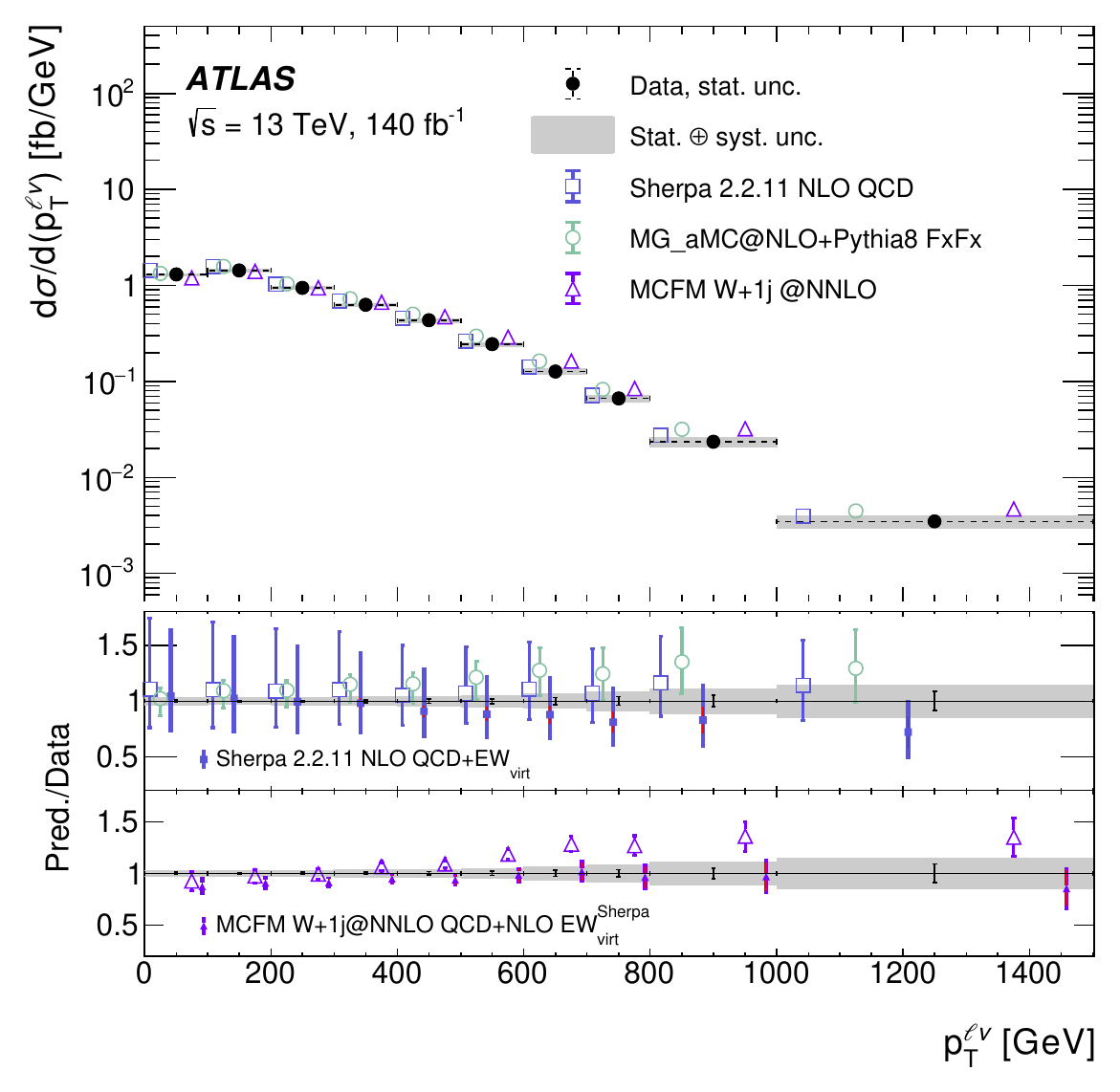} \label{fig:result:wpt}  }
\caption{
Differential \xsec as a function of (a) the leading \ptjet, (b) the inclusive jet multiplicity, (c) \HT, and (d) \pTW in the collinear phase-space.
The measured \xsecs in data and their statistical uncertainty are shown by the solid dots and error bars,
while the shaded band shows the systematic and statistical uncertainties added in quadrature.
Errors bars on the theory prediction include theoretical uncertainties as discussed in the text.
The two lower panels show the effect of NLO EW$_{\text{virt}}$ as computed from \SHERPA[2.2.11].
}
\label{fig:unfolding:unfolded:differential:collinear}
\end{center}
\end{figure}

\FloatBarrier



\section{Conclusions}
\label{sec:conclusions}

A measurement of \xsecs for a \Wboson produced in association with at least one high-transverse-momentum jet is presented.
This measurement utilises 140~\ifb{} of proton--proton collision data collected at a centre-of-mass energy of $\sqrt{s}$~=~13~\TeV by the ATLAS detector at the LHC.
Measurements are performed on events containing a single electron or muon from the $W\rightarrow \ell\nu$ decay
and at least one high-momentum jet with \pt{}$>$~500~\GeV.
This paper focuses on events where the angular separation between the lepton
and a high-momentum jet is small to define a collinear phase space.
This region is populated either by dijet events radiating a \Wboson or events with a \Wboson produced in association
with several jets and it serves as an excellent data sample to probe higher-order theoretical predictions.
Measurements of the inclusive and differential \xsecs in the collinear phase space
as a function of a variety of observables sensitive to the emission of \Wbosons from high-momentum jets are presented.
Measurements probe \Wbosons produced in association with jets with transverse momentum above the \TeV scale.

The background-subtracted data distributions are unfolded to the particle level
and compared with several multijet generator predictions and to a fixed-order calculation.
The \SHERPA and \mgamc{} +\pythia multijet merged predictions, accurate to NLO in $\alphas$,
provide an excellent description of the data across all measured distributions.
The scale uncertainties in the \SHERPA predictions are found to be significantly larger than those from \mgamc,
due to the extra leading-order matrix elements in the \SHERPA sample used to describe the high-jet-multiplicity final states.
The fixed-order calculation for \Wplusonejet from MCFM, computed to NNLO in $\alphas$,
provides a good description of the data with a precision that is comparable to the measurement uncertainties.
In regions of phase space with large angular separations between the lepton and leading jet, the MCFM prediction underestimates the \xsec.
The impact of NLO EW virtual corrections is assessed using the \SHERPA event generator,
the size of the corrections is found to be larger than the scale uncertainties in the MCFM prediction.
NLO EW corrections improve the agreement of the theory prediction with measured data
in regions of phase space with highly-boosted \Wbosons.


%

\section*{Acknowledgements}
%

%
%

%
%

We thank CERN for the very successful operation of the LHC and its injectors, as well as the support staff at
CERN and at our institutions worldwide without whom ATLAS could not be operated efficiently.

The crucial computing support from all WLCG partners is acknowledged gratefully, in particular from CERN, the ATLAS Tier-1 facilities at TRIUMF/SFU (Canada), NDGF (Denmark, Norway, Sweden), CC-IN2P3 (France), KIT/GridKA (Germany), INFN-CNAF (Italy), NL-T1 (Netherlands), PIC (Spain), RAL (UK) and BNL (USA), the Tier-2 facilities worldwide and large non-WLCG resource providers. Major contributors of computing resources are listed in Ref.~\cite{ATL-SOFT-PUB-2025-001}.

We gratefully acknowledge the support of ANPCyT, Argentina; YerPhI, Armenia; ARC, Australia; BMWFW and FWF, Austria; ANAS, Azerbaijan; CNPq and FAPESP, Brazil; NSERC, NRC and CFI, Canada; CERN; ANID, Chile; CAS, MOST and NSFC, China; Minciencias, Colombia; MEYS CR, Czech Republic; DNRF and DNSRC, Denmark; IN2P3-CNRS and CEA-DRF/IRFU, France; SRNSFG, Georgia; BMBF, HGF and MPG, Germany; GSRI, Greece; RGC and Hong Kong SAR, China; ICHEP and Academy of Sciences and Humanities, Israel; INFN, Italy; MEXT and JSPS, Japan; CNRST, Morocco; NWO, Netherlands; RCN, Norway; MNiSW, Poland; FCT, Portugal; MNE/IFA, Romania; MSTDI, Serbia; MSSR, Slovakia; ARIS and MVZI, Slovenia; DSI/NRF, South Africa; MICIU/AEI, Spain; SRC and Wallenberg Foundation, Sweden; SERI, SNSF and Cantons of Bern and Geneva, Switzerland; NSTC, Taipei; TENMAK, T\"urkiye; STFC/UKRI, United Kingdom; DOE and NSF, United States of America.

Individual groups and members have received support from BCKDF, CANARIE, CRC and DRAC, Canada; CERN-CZ, FORTE and PRIMUS, Czech Republic; COST, ERC, ERDF, Horizon 2020, ICSC-NextGenerationEU and Marie Sk{\l}odowska-Curie Actions, European Union; Investissements d'Avenir Labex, Investissements d'Avenir Idex and ANR, France; DFG and AvH Foundation, Germany; Herakleitos, Thales and Aristeia programmes co-financed by EU-ESF and the Greek NSRF, Greece; BSF-NSF and MINERVA, Israel; NCN and NAWA, Poland; La Caixa Banking Foundation, CERCA Programme Generalitat de Catalunya and PROMETEO and GenT Programmes Generalitat Valenciana, Spain; G\"{o}ran Gustafssons Stiftelse, Sweden; The Royal Society and Leverhulme Trust, United Kingdom.

In addition, individual members wish to acknowledge support from Armenia: Yerevan Physics Institute (FAPERJ); CERN: European Organization for Nuclear Research (CERN DOCT); Chile: Agencia Nacional de Investigaci\'on y Desarrollo (FONDECYT 1230812, FONDECYT 1230987, FONDECYT 1240864); China: Chinese Ministry of Science and Technology (MOST-2023YFA1605700, MOST-2023YFA1609300), National Natural Science Foundation of China (NSFC - 12175119, NSFC 12275265, NSFC-12075060); Czech Republic: Czech Science Foundation (GACR - 24-11373S), Ministry of Education Youth and Sports (FORTE CZ.02.01.01/00/22\_008/0004632), PRIMUS Research Programme (PRIMUS/21/SCI/017); EU: H2020 European Research Council (ERC - 101002463); European Union: European Research Council (ERC - 948254, ERC 101089007, ERC, BARD, 101116429), European Union, Future Artificial Intelligence Research (FAIR-NextGenerationEU PE00000013), Italian Center for High Performance Computing, Big Data and Quantum Computing (ICSC, NextGenerationEU); France: Agence Nationale de la Recherche (ANR-20-CE31-0013, ANR-21-CE31-0013, ANR-21-CE31-0022, ANR-22-EDIR-0002); Germany: Baden-Württemberg Stiftung (BW Stiftung-Postdoc Eliteprogramme), Deutsche Forschungsgemeinschaft (DFG - 469666862, DFG - CR 312/5-2); Italy: Istituto Nazionale di Fisica Nucleare (ICSC, NextGenerationEU), Ministero dell'Università e della Ricerca (PRIN - 20223N7F8K - PNRR M4.C2.1.1); Japan: Japan Society for the Promotion of Science (JSPS KAKENHI JP22H01227, JSPS KAKENHI JP22H04944, JSPS KAKENHI JP22KK0227, JSPS KAKENHI JP23KK0245); Netherlands: Netherlands Organisation for Scientific Research (NWO Veni 2020 - VI.Veni.202.179); Norway: Research Council of Norway (RCN-314472); Poland: Ministry of Science and Higher Education (IDUB AGH, POB8, D4 no 9722), Polish National Agency for Academic Exchange (PPN/PPO/2020/1/00002/U/00001), Polish National Science Centre (NCN 2021/42/E/ST2/00350, NCN OPUS 2023/51/B/ST2/02507, NCN OPUS nr 2022/47/B/ST2/03059, NCN UMO-2019/34/E/ST2/00393, NCN \& H2020 MSCA 945339, UMO-2020/37/B/ST2/01043, UMO-2021/40/C/ST2/00187, UMO-2022/47/O/ST2/00148, UMO-2023/49/B/ST2/04085, UMO-2023/51/B/ST2/00920); Slovenia: Slovenian Research Agency (ARIS grant J1-3010); Spain: Generalitat Valenciana (Artemisa, FEDER, IDIFEDER/2018/048), Ministry of Science and Innovation (MCIN \& NextGenEU PCI2022-135018-2, MICIN \& FEDER PID2021-125273NB, RYC2019-028510-I, RYC2020-030254-I, RYC2021-031273-I, RYC2022-038164-I); Sweden: Carl Trygger Foundation (Carl Trygger Foundation CTS 22:2312), Swedish Research Council (Swedish Research Council 2023-04654, VR 2018-00482, VR 2021-03651, VR 2022-03845, VR 2022-04683, VR 2023-03403), Knut and Alice Wallenberg Foundation (KAW 2018.0458, KAW 2019.0447, KAW 2022.0358); Switzerland: Swiss National Science Foundation (SNSF - PCEFP2\_194658); United Kingdom: Leverhulme Trust (Leverhulme Trust RPG-2020-004), Royal Society (NIF-R1-231091); United States of America: U.S. Department of Energy (ECA DE-AC02-76SF00515), Neubauer Family Foundation.

%
%


%
%
%
\clearpage
\printbibliography
\clearpage
 
\begin{flushleft}
\hypersetup{urlcolor=black}
{\Large The ATLAS Collaboration}

\bigskip

\AtlasOrcid[0000-0002-6665-4934]{G.~Aad}$^\textrm{\scriptsize 104}$,
\AtlasOrcid[0000-0001-7616-1554]{E.~Aakvaag}$^\textrm{\scriptsize 17}$,
\AtlasOrcid[0000-0002-5888-2734]{B.~Abbott}$^\textrm{\scriptsize 123}$,
\AtlasOrcid[0000-0002-0287-5869]{S.~Abdelhameed}$^\textrm{\scriptsize 119a}$,
\AtlasOrcid[0000-0002-1002-1652]{K.~Abeling}$^\textrm{\scriptsize 56}$,
\AtlasOrcid[0000-0001-5763-2760]{N.J.~Abicht}$^\textrm{\scriptsize 50}$,
\AtlasOrcid[0000-0002-8496-9294]{S.H.~Abidi}$^\textrm{\scriptsize 30}$,
\AtlasOrcid[0009-0003-6578-220X]{M.~Aboelela}$^\textrm{\scriptsize 46}$,
\AtlasOrcid[0000-0002-9987-2292]{A.~Aboulhorma}$^\textrm{\scriptsize 36e}$,
\AtlasOrcid[0000-0001-5329-6640]{H.~Abramowicz}$^\textrm{\scriptsize 155}$
% ...
\bigskip
\\

% ...
$^{17}$Department for Physics and Technology, University of Bergen, Bergen; Norway.\\
% ...
$^{30}$Physics Department, Brookhaven National Laboratory, Upton NY; United States of America.\\
% ...
$^{36}$$^{(a)}$Facult\'e des Sciences Ain Chock, Universit\'e Hassan II de Casablanca;$^{(b)}$Facult\'{e} des Sciences, Universit\'{e} Ibn-Tofail, K\'{e}nitra;$^{(c)}$Facult\'e des Sciences Semlalia, Universit\'e Cadi Ayyad, LPHEA-Marrakech;$^{(d)}$LPMR, Facult\'e des Sciences, Universit\'e Mohamed Premier, Oujda;$^{(e)}$Facult\'e des sciences, Universit\'e Mohammed V, Rabat;$^{(f)}$Institute of Applied Physics, Mohammed VI Polytechnic University, Ben Guerir; Morocco.\\
% ...
$^{46}$Physics Department, Southern Methodist University, Dallas TX; United States of America.\\
% ...
$^{50}$Fakult\"{a}t Physik , Technische Universit{\"a}t Dortmund, Dortmund; Germany.\\
% ...
$^{56}$II. Physikalisches Institut, Georg-August-Universit\"{a}t G\"ottingen, G\"ottingen; Germany.\\
% ...
$^{104}$CPPM, Aix-Marseille Universit\'e, CNRS/IN2P3, Marseille; France.\\
% ...
$^{119}$$^{(a)}$New York University Abu Dhabi, Abu Dhabi;$^{(b)}$United Arab Emirates University, Al Ain; United Arab Emirates.\\
% ...
$^{123}$Homer L. Dodge Department of Physics and Astronomy, University of Oklahoma, Norman OK; United States of America.\\
% ...
$^{155}$Raymond and Beverly Sackler School of Physics and Astronomy, Tel Aviv University, Tel Aviv; Israel.\\
% ...
$^{a}$ Also Affiliated with an institute covered by a cooperation agreement with CERN.\\
$^{b}$ Also at An-Najah National University, Nablus; Palestine.\\
$^{c}$ Also at Borough of Manhattan Community College, City University of New York, New York NY; United States of America.\\
$^{d}$ Also at Center for High Energy Physics, Peking University; China.\\
$^{e}$ Also at Center for Interdisciplinary Research and Innovation (CIRI-AUTH), Thessaloniki; Greece.\\
$^{f}$ Also at CERN, Geneva; Switzerland.\\
% ...

\end{flushleft}


%

%
%

%

%
%
%
%

\end{document}